\documentclass[twocolumn]{aastex63}

\usepackage{verbatim} 
\usepackage{amsmath}

\usepackage{float}

\submitjournal{ApJ}
\shorttitle{Optical IFU observations of the dual AGN Mrk 739}
\shortauthors{Tub\'in et al.}

\begin{document}

\title{The Complex Gaseous and Stellar environments of the nearby dual AGN Mrk 739} 

\correspondingauthor{Dus\'an Tub\'in}
\email{dtubin@astro.puc.cl}

\author[0000-0002-2688-7960]{Dus\'an Tub\'in}
\affil{Instituto de Astrof{\'i}sica, Facultad de F{\'i}sica, Pontificia Universidad Cat{\'o}lica de Chile, Casilla 306, Santiago 22, Chile}

\author[0000-0001-7568-6412]{Ezequiel Treister}
\affiliation{Instituto de Astrof{\'i}sica, Facultad de F{\'i}sica, Pontificia Universidad Cat{\'o}lica de Chile, Casilla 306, Santiago 22, Chile}

\author[0000-0001-9697-7331]{Giuseppe D'Ago}
\affiliation{Instituto de Astrof{\'i}sica, Facultad de F{\'i}sica, Pontificia Universidad Cat{\'o}lica de Chile, Casilla 306, Santiago 22, Chile}

\author[0000-0001-8349-3055]{Giacomo Venturi}
\affiliation{Instituto de Astrof{\'i}sica, Facultad de F{\'i}sica, Pontificia Universidad Cat{\'o}lica de Chile, Casilla 306, Santiago 22, Chile}
\affiliation{INAF - Osservatorio Astrofisico di Arcetri, Largo E. Fermi 5, I-50125 Firenze, Italy}

\author[0000-0002-8686-8737]{Franz E. Bauer}
\affiliation{Instituto de Astrof{\'{\i}}sica and Centro de Astroingenier{\'{\i}}a, Facultad de F{\'{i}}sica, Pontificia Universidad Cat{\'{o}}lica de Chile, Casilla 306, Santiago 22, Chile} 
\affiliation{Millennium Institute of Astrophysics (MAS), Nuncio Monse{\~{n}}or S{\'{o}}tero Sanz 100, Providencia, Santiago, Chile} 
\affiliation{Space Science Institute, 4750 Walnut Street, Suite 205, Boulder, Colorado 80301, USA} 

\author[0000-0003-3474-1125]{George C. Privon}
\affiliation{National Radio Astronomy Observatory, 520 Edgemont Rd, Charlottesville,
VA 22903, USA}

\author[0000-0002-7998-9581]{Michael J. Koss}
\affiliation{Eureka Scientific, 2452 Delmer Street Suite 100, Oakland, CA 94602-3017, USA}
\affiliation{Space Science Institute, 4750 Walnut Street, Suite 205, Boulder, Colorado 80301, USA}

\author[0000-0001-5742-5980]{Federica Ricci}
\affiliation{Dipartimento di Fisica e Astronomia, Universit\`a di Bologna, via Gobetti 93/2, 40129 Bologna, Italy}
\affiliation{INAF- Osservatorio di Astrofisica e Scienza dello Spazio di Bologna, via Gobetti 93/3, 40129 Bologna, Italy} 

\author{Julia M. Comerford}
\affiliation{Department of Astrophysical and Planetary Sciences, University of Colorado, Boulder, CO 80309, USA}

\author[0000-0002-2713-0628]{Francisco M\"{u}ller-S\'anchez}
\affiliation{Department of Physics and Materials Science, The University of Memphis, 3720 Alumni Ave, Memphis, TN 38152}

\begin{abstract}
We present Integral Field Spectroscopic (IFS) observations of the nearby ($z\sim0.03$) dual Active Galactic Nuclei (AGN) Mrk 739, whose projected nuclear separation is $\sim$3.4~kpc, obtained with the Multi Unit Spectroscopic Explorer (MUSE) at the Very Large Telescope (VLT). We find that the galaxy has an extended AGN-ionized emission-line region extending up to $\sim 20$ kpc away from the nuclei, while star-forming regions are more centrally concentrated within 2 - 3 kpc. We model the kinematics of the ionized gas surrounding the East nucleus using a circular disk profile, resulting in a peak velocity of $237^{+26}_{-28}$ km s$^{-1}$ at a distance of $\sim 1.2$ kpc. The enclosed dynamical mass within 1.2 kpc is $\log M(M_{\odot})=10.20\pm0.06$, $\sim$1,000 times larger than the estimated supermassive black hole (SMBH) mass of Mrk 739E. The morphology and dynamics of the system are consistent with an early stage of the collision, where the foreground galaxy (Mrk 739W) is a young star-forming galaxy in an ongoing first passage with its background companion (Mrk 739E). Since the SMBH in Mrk 739W does not show evidence of being rapidly accreting, we claim that the northern spiral arms of Mrk 739W are ionized by the nuclear activity of Mrk 739E.

\end{abstract}


\section{Introduction}\label{sec:intro}

Nuclear activity, produced by an active galactic nucleus (AGN), which is powered by accretion on a supermassive black hole (SMBH), and major galaxy mergers are key processes to understand the formation and evolution of galaxies \citep[e.g.,][]{1988ApJ...325...74S}. The effects of the interaction between the central nuclear engine and the rest of the galaxy can play a fundamental role in the evolution of the galaxy \citep{1998A&A...331L...1S}. Correlations spanning several orders of magnitude exist between the mass of the central SMBH and the properties of the host galaxy \citep{2000ApJ...539L...9F,2000ApJ...539L..13G,2013ARA&A..51..511K}, e.g., the SMBH mass and the velocity dispersion of bulge $M_{\rm BH}-\sigma$ \citep{Greene_2006,2009ApJ...698..198G}, the mass of its spheroidal component \citep{1998AJ....115.2285M} and the bulge luminosity \citep{2003ApJ...589L..21M,2007MNRAS.379..711G}, respectively. These relations can be explained by a connection between the SMBH growth and the available gas, suggesting that the fueling of the host gas reservoir on the SMBH is regulated by AGN activity in the form of energetic radiation, outflows, and jets \citep[section 2.7]{2012RAA....12..917S}. This energetic AGN ``feedback'' affects the interstellar medium of the host galaxy \citep{2013ARA&A..51..511K,2012ARA&A..50..455F}, either igniting \citep{Shin_2019} or suppressing \citep{Alatalo_2014,2016Natur.533..504C}  star formation. Hence, it is clear that nuclear activity is a critical ingredient for galaxy evolution, albeit not the only one. 

Major galaxy interactions are important events in order to understand the growth of the SMBH \citep{2005Natur.433..604D} as well as the process and history of star formation \citep{1988ApJ...325...74S}. \citet{2006ApJS..163....1H} synthesized the formation and evolution of the galaxies with a ``cosmic cycle''. In this model, activity directly linked to a galaxy merger can drive the gas towards the nucleus due to gravitational torques \citep{1996ApJ...464..641M}. This gas inflow can then trigger star formation and fuel SMBH growth, hence causing the so-called ``quasar'' (luminous AGN) phase. The high gas density obscures the source until the energy released by the AGN expels the gas, making the quasar visible \citep{Treister600}. The energy of the outflow is enough to remove the gas and dust that feed the SMBH and to quench further star formation and black hole growth, leaving as a remnant a more massive black hole and a stellar spheroidal component.

The enhancement of star formation in interacting systems is a direct consequence of the gravitational forces and tidal disruptions produced in galaxy mergers \citep{2004MNRAS.355..874N,2011MNRAS.412..591P,2012A&A...548A.117Y}. In most nearby strongly star-forming galaxies, the dominant trigger of star formation is attributed to tidal interactions \citep{Li_2008}. However, it is not always the case that the global star formation of merging systems is significantly higher than in isolated galaxies \citep{2003A&A...405...31B}. In a recent study, \citet{Pearson_2019} concluded that the star formation rates (SFRs) of galaxy mergers are not significantly different from those of non-interacting galaxies. However, the higher the SFR, the higher is the fraction of merging galaxies, thus confirming that indeed galaxy mergers can induce strong star formation episodes.

During the merging process, when the two nuclei are closer than 10 kpc, and both SMBHs are actively accreting the surrounding material, the system is considered a dual AGN. Hydrodynamical simulations \citep{Van_Wassenhove_2012,2018arXiv180501479R} show that the fraction of detectable dual AGN increases with decreasing separation between the nuclei ($<$1-10 kpc). This is consistent with observations that show that the X-ray luminosity of the system increases with decreasing SMBHs separation \citep{2012ApJ...746L..22K}, strengthening the idea that galaxy merging can act as a  trigger for nuclear activity. 

At optical wavelengths, at $0.02< z < 0.16$, the observed fraction of dual AGN among all spectroscopically selected AGNs, as  reported by \cite{Liu_2011}, is $\sim 3.6\%$. This low fraction can be explained by the strong obscuration that the optical wavelengths can experience at different times of the merger. Observationally, a lower [OIII] to X-ray has been found in mergers \citep{koss10}, together with higher fractions of obscured AGN at smaller nuclear separations \citep{2018Natur.563..214K}. Indeed, at the most advanced stages of the collision, a link between merger fraction and obscuration has been previously found \citep{2015ApJ...814..104K,2016ApJ...825...85K}. It is expected that the gas and dust of the galaxy can obscure up to $\sim 95\%$ of the central X-ray source \citep{2017MNRAS.468.1273R}.

The characterization of local, confirmed, dual AGN at different evolutionary and morphological stages helps us to understand the physical properties and kinematics across the merger sequence. Notable examples include Mrk 463, cataloged as a dual AGN by  \cite{2008MNRAS.386..105B} and recently characterized by \cite{2018ApJ...854...83T}. In that work, multi-wavelength data of Mrk 463 provide strong evidence for a biconical outflow with velocities $>600$ km s$^{-1}$, associated with the Mrk 463E nucleus. Moreover, using one of the most advanced techniques of Adaptive Optic (AO) at the Multi-Unit Spectroscopic Explorer \citep[MUSE,][]{2010SPIE.7735E..08B} instrument at the Very Large Telescope (VLT), \citet{2020A&A...633A..79K} present the discovery of a rare triple AGN candidate in the NGC 6240 galaxy.  

In this paper, we expand the characterization of local AGN pairs, with a detailed study of the dual AGN Mrk 739 observed with the optical and near-IR MUSE instrument at the VLT. This is a nearby ($z$=0.02985 or $d$ $\sim$130 Mpc) interacting system with a projected nuclear separation of 3.4 kpc \citep{koss10}. It was later classified as a dual AGN by \citet{2011ApJ...735L..42K} based on \textit{Chandra} X-ray observations and is one of the 17 sources studied by the Multiwavelength Observations of dual AGN (MODA\footnote{\url{http://moda.astro.puc.cl/}}) project.

Mrk 739 is a particularly interesting source. The western nucleus was not classified as an AGN based on either optical emission-line diagnostics or radio and UV observations. It could most likely be explained by the presence of a strong HII region \citep{1987A&A...171...41N}, causing the low $L_{\rm [OIII]}/L_{2-10 \rm keV}$ ratio observed in the nuclear region. Based on \textit{Chandra} data, \cite{2011ApJ...735L..42K} found in the central region of Mrk 739 two X-ray sources. The East nucleus, Mrk 739E, was found to have an absorption-corrected luminosity of $L_{2-10 \rm keV}=1.1\times10^{43}$ erg s$^{-1}$ and a column density $N_{\rm H}=1.5 \pm 0.2 \times 10^{21}$ cm$^{-2}$, consistent with its optical spectral classification as a Seyfert 1 galaxy. The West nucleus, Mrk 739W, has an absorption-corrected luminosity of $L_{2-10 \rm keV}=1.0\times10^{42}$ erg s$^{-1}$ and a slightly higher absorption with $N_{H}=4.6 \pm 0.1 \times 10^{21}$ cm$^{-2}$. Similar to Mrk 463 and NGC 6240, Mrk 739 is a bright hard X-ray source ($L_{14-195 \rm keV}=2.4\times 10^{43}\rm erg\;\rm s^{-1}$) and it was observed as part of routine follow-up of mergers  \citep{koss10} in the Swift BAT sample, which observes the brightest AGN in the sky at 14-195 keV \citep{2013ApJS..207...19B}.

In this paper, we present VLT/MUSE IFU spectroscopic data of the galaxy merging system Mrk 739 in order to determine and understand the nature of the stellar population, in terms of age and metallicity. We also study the ionized gas, traced by several emission lines, in terms of morphology, excitation mechanism, and velocity. By investigating the behavior of the ionized gas and the stellar population, we aim to characterize the morphology of the system and understand the effects of the interaction on the host galaxy. This paper is organized as follows: We present in Section 2 the properties of the IFS data and the analysis of the spectral cube. In section 3, we detail the morphology of the main emission lines in the spectral range covered by MUSE, the nature of the ionization source, and the kinematics of the ionized gas traced by emission lines. The results of the stellar populations fitting are presented in Section 4, while sections 5 and 6 report discussion and conclusions, respectively.  Throughout this paper, we assume a $\Lambda$CDM cosmology with $h_{0}=0.7$,
$\Omega_{m}=0.27$ and $\Omega_{\Lambda}=0.73$ \citep{2009ApJS..180..225H}. In all the figures, the standard astronomical orientation with North up and East to the left was adopted.

\section{Data description and Analysis}\label{sec:data}

\begin{figure*}
\centering
\includegraphics[width=0.8\textwidth,keepaspectratio=true]{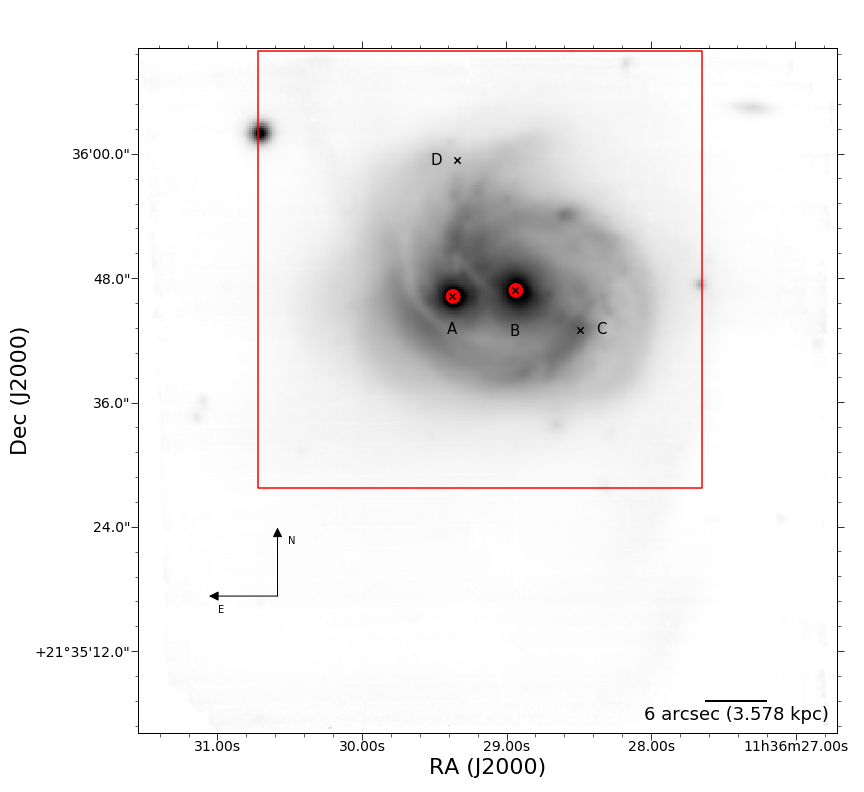}
\caption{Reconstructed VLT/MUSE white-light image of the major galaxy merger Mrk 739, covering the optical wavelength range between 4800-9300\AA. The seeing-limited spatial resolution is $\sim0\farcs7$. Red circles show the location of the X-ray emission associated with the two nuclei, as obtained from the Second Chandra X-ray Source Catalog (CSC 2.0) with a 95\% confidence level positional error of $0\farcs71$ \citep{evans20}.  The black crosses mark the location of the extracted spectra shown in Figure \ref{fig:selectedspectra}. The red square marks the field of view presented in the subsequent maps.}
\label{fig:whiteimage}
\end{figure*}

Mrk 739 was observed with the VLT/MUSE IFU spectrograph in the no-AO Wide Field Mode (WFM), as part of ESO program 095.B-0482 (PI: E. Treister). MUSE is a second-generation VLT instrument that operates in the optical and near-IR spectral range at wavelengths between 4800 and 9300 \AA, and in the local Universe, it covers spectral emission lines such as H$\beta$ $\lambda4861$, H$\alpha$ $\lambda6563$ and [OIII] $\lambda5007$, with a field of view (FoV) of 1 arcmin$^{2}$. The observations were carried out in service mode, with clear sky conditions, $<50\%$ lunar illumination, and an average seeing of $0\farcs76$.
With a total exposure time of 5856 seconds ($\sim 1.6$ hours), the final reduced cube is a combination of 6 raw science exposures. The calibration and data reduction were carried out using the ESO VLT/MUSE pipeline \citep{2014ASPC..485..451W} with the standard instrumental corrections (bias, dark, flat fielding corrections, flux, and wavelength calibrations) under the ESO \textit{Reflex} environment \citep{2013A&A...559A..96F}. In Figure \ref{fig:whiteimage}, we present the white-light image of the MUSE cube collapsed along the spectral axis and the positions of the X-ray emission associated with both nuclei with the corresponding $95\%$ confidence level of $0\farcs71$ represented by the red circles.

\begin{figure*}
\centering
\includegraphics[width=1.0\textwidth,keepaspectratio=true]{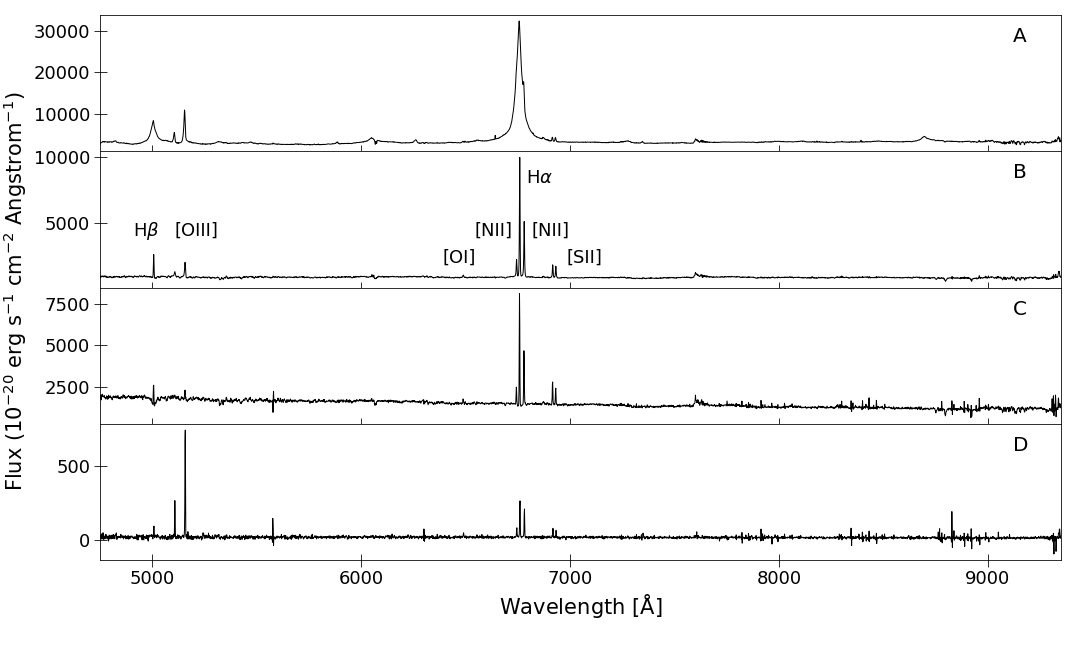}
\caption{ Representative VLT/MUSE single-pixel spectra extracted from the regions presented in Figure \ref{fig:whiteimage}. Spectrum A shows broad Balmer emission lines originating from the nuclear region of Mrk 739E. The H$\alpha$-[NII] complex is difficult to separate due to the strong contribution of the broad emission. Spectrum B presents the narrow emission lines of Mrk 739W, without evidence of broad components. Spectrum C exhibits narrow emission lines, together with stellar features such as a Balmer absorption line in H$\beta$. Spectrum D shows the fainter emission representative from the northern regions of the FoV.}
\label{fig:selectedspectra}
\end{figure*}

A quick look at the data cube was performed using QFitsView\footnote{\url{http://www.mpe.mpg.de/\~ott/QFitsView/}}.  Figure \ref{fig:selectedspectra} shows representative VLT/MUSE spectra of four distinct regions, covering the entire instrumental wavelength range. We clearly identify broad Balmer emission lines on Spectrum A, in the nuclear region of Mrk 739E. On the other hand, the B spectrum, characterizing the emission from Mrk 739W, does not show evidence of broad lines, as previously reported by \cite{2011ApJ...735L..42K}. Spectra C and D show two representative regions from the host galaxy. Spectrum C covers the emission from a spiral structure located in the south-west direction, where the [OIII]$\lambda5007$ is faint, and we can find overlapping H$\beta$ emission and absorption, typically associated with post-starburst episodes \citep{2007MNRAS.381..187G,2012MNRAS.420.1684W}, likely produced by the gas-rich major merger \citep{2018MNRAS.477.1708P}. The D spectrum characterizes the extended emission in the northern regions, where the emission is faint, and the predominant line is [OIII]$\lambda5007$.

In order to correct for absorption and provide a better estimation of the flux of the overlapping emission lines, we follow a similar procedure to \cite{2018A&A...619A..74V} to subtract the continuum emission. Basically, the method applies a Voronoi tessellation \citep{2003MNRAS.342..345C} to the full cube to guarantee a minimum signal-to-noise ratio (SNR) of 40 per wavelength channel in the full observe-frame spectral window range. We considered as input for the Voronoi tessellation the spaxels with an individual SNR $>1$. The binned spectra were then fitted using the pPXF \citep{Cappellari2017} package to derive the kinematic and parameters like age and metallicity of the stellar content. Briefly, our fitting procedure masks the most prominent emission lines of ionized gas and sky, and fits the stellar continuum using templates of single stellar populations (SSP) from the extended MILES (E-MILES) library \citep{2016MNRAS.463.3409V}. A more detailed explanation of the stellar population analysis is later presented in \S \ref{stellarpopulation}, where we also show the main physical properties of the stellar components. 

\begin{figure*}
\centering
\includegraphics[width=1.0\textwidth,height=11cm,keepaspectratio=true]{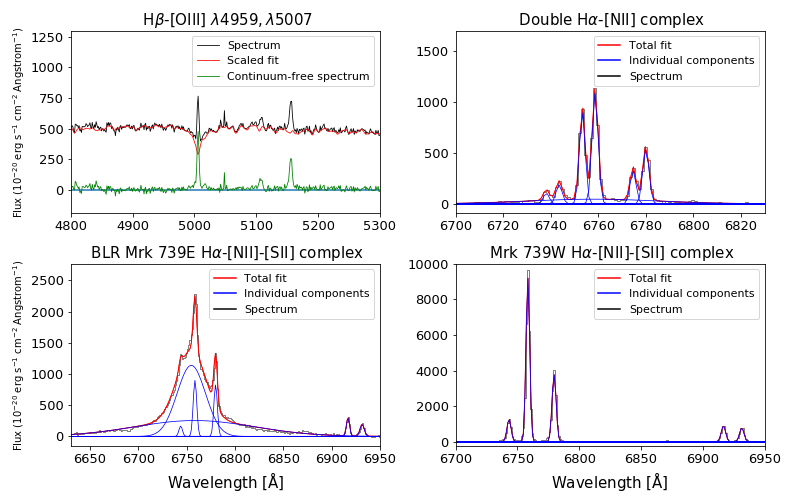}
\caption{Representative spectral fits from extracted regions. {\it Top panels}: A Balmer absorption region ({\it left}), where the continuum-free emission (\textit{green curve}) is the result of the difference between the observed spectrum (\textit{black curve}) and the flux-scaled best-fit of the continuum (\textit{red curve}). This fit highlights the importance of accurately modelling the continuum template to recover all of the H$\beta$ flux. The double-peaked emission line region ({\it right}) is representative of the emission from the right side of Mrk 739E and is fit with two narrow components (\textit{blue curves}) per emission line. {\it Lower panels}: The broad line region in the nuclear portion of Mrk 739E ({\it left}) is fit with two broad Gaussian components for each Balmer line, and one narrow component that describes the diffuse emission. The emission lines of Mrk 739W ({\it right}) are fit with narrow Gaussian components since broad Balmer lines are obscured by star formation. The total modeled emission is displayed in red for the {\it top-right}, {\it lower-left} and {\it lower-right} panels.}
\label{fig:representativespectra}
\end{figure*}

Then, for every spaxel in the unbinned data cube that is associated with a given bin, we subtracted a scaled version of the best-fit stellar continuum. Since we masked the most prominent emission lines and therefore also part of the absorption lines, we use the full-spectrum fitting method to recover the Balmer absorption features from the stellar population templates that give the best-fit to the rest of the continuum. A graphical representation of this method is presented on the \textit{top-left} panel of Figure \ref{fig:representativespectra}. It should be noted that our resulting continuum-free cube, which contains the absorption-corrected emission lines, only gives a model-dependent estimate of the total flux of the emission lines. In order to incorporate those individual spaxels below the SNR $>1$ cut in the continuum but with well-defined emission lines, or those that were discarded by the continuum fitting due to the presence of a broad component (see \S\ref{stellarpopulation}), we fitted and subtracted a second-order polynomial function to unbinned spaxels, excluding the regions where there are emission lines, which are hence unaffected by the subtraction. 

Finally, since the H$\beta$ Balmer line is one of the weakest lines in this system and the instrument efficiency and spectral resolution are lower at blue wavelengths, we performed a second Voronoi binning on the emission-line cube, now requiring a SNR $>$5 per bin around H$\beta$, in the observed wavelength range between 5000-5010 \AA. The atomic features covered in the MUSE wavelength range were then fitted using Pyspeckit, the Python Spectroscopic Toolkit package \citep{2011ascl.soft09001G}. Given the redshift of Mrk 739 and the wavelength coverage of MUSE, the following emission lines are observed: H$\beta$ $\lambda4861$, [OIII] $\lambda4959,\lambda5007$, [OI] $\lambda6300$ H$\alpha\;\lambda6563$ [NII] $\lambda6549,\lambda6583$ and [SII] $\lambda6717,\lambda6730$. A set of 2 narrows ($\sigma < 500$ km s$^{-1}$) Gaussian components were used to fit each of the mentioned emission lines in each binned spectrum. We found that, for a given emission line, one narrow component is capable of reproducing the emission of the whole galaxy, while the second component is required to account for a very particular region close to the East AGN. An example spectrum from this region is shown on the \textit{top-right} panel of Figure \ref{fig:representativespectra}. We found that this extra emission component (hereafter called ``second component'') presents a blueshifted velocity offset with respect to the systemic component that fits the emission of the rest of the galaxy. Further details about the kinematics of this region are presented on \S \ref{doubleline}. For better visualization, in the top-right panel of Figure \ref{fig:halphahbetaoiiioi}, we highlight with yellow contours the region where the second emission line component is found. For clarity, we only report the second component of this region in the following ionized gas maps.

Additionally, to reproduce the Broad Line Region (BLR) emission of the Balmer lines, we incorporated two broad ($300<\sigma < 3300$ km s$^{-1}$) components for H$\beta$ and two for H$\alpha$ (\textit{lower-left} panel in Figure \ref{fig:representativespectra}). We fixed the ratios of the [OIII] and [NII] doublet to their theoretically-determined values of 3 \citep{2006agna.book.....O} in order to reduce the degrees of freedom of the fit. Given that it is a strong and well-isolated feature, we use the [OIII]$\lambda5007$ line as a template in order to define the widths for all the other narrow components. This choice does not leave significant residuals in the other lines. Since the MUSE spectral resolution is close to $R\sim 2500$ at $\sim 6550$\AA, we can separate the H$\alpha$ component from the [NII] lines accurately across nearly the entire cube. The H$\alpha$-[NII] complex is difficult to constrain only when the BLR contributes with velocities larger than 500 km s$^{-1}$, which occurs near the Mrk 739E nucleus.

\begin{figure*}[ht]
\begin{minipage}[t]{0.49\linewidth}
\includegraphics[width=1.\linewidth,keepaspectratio=True]{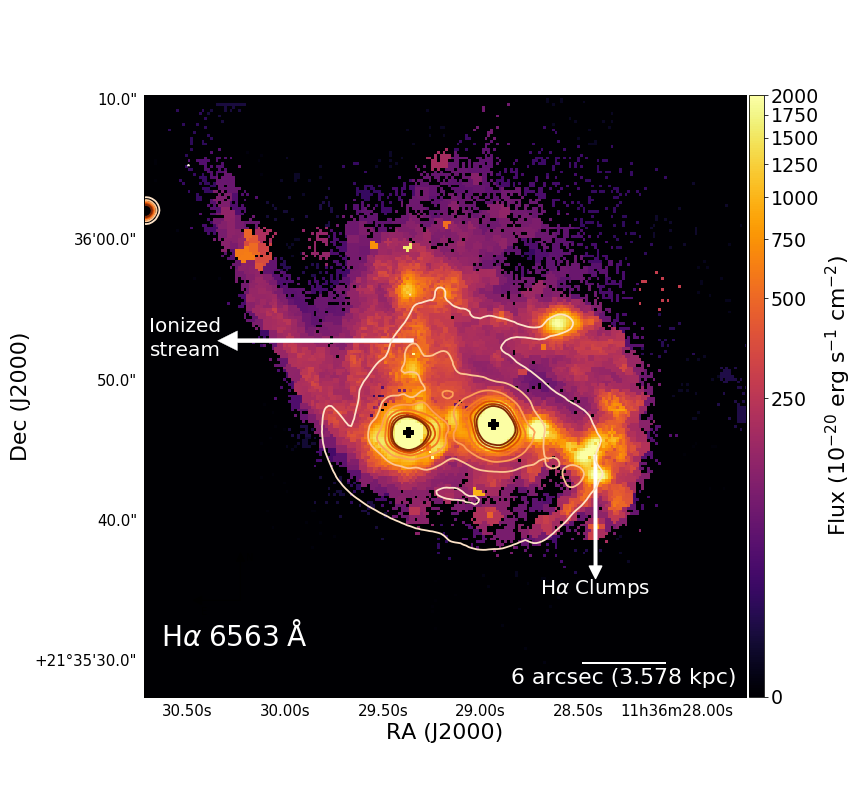}
\end{minipage}
\begin{minipage}[t]{0.49\linewidth}
\includegraphics[width=1.\linewidth,keepaspectratio=True]{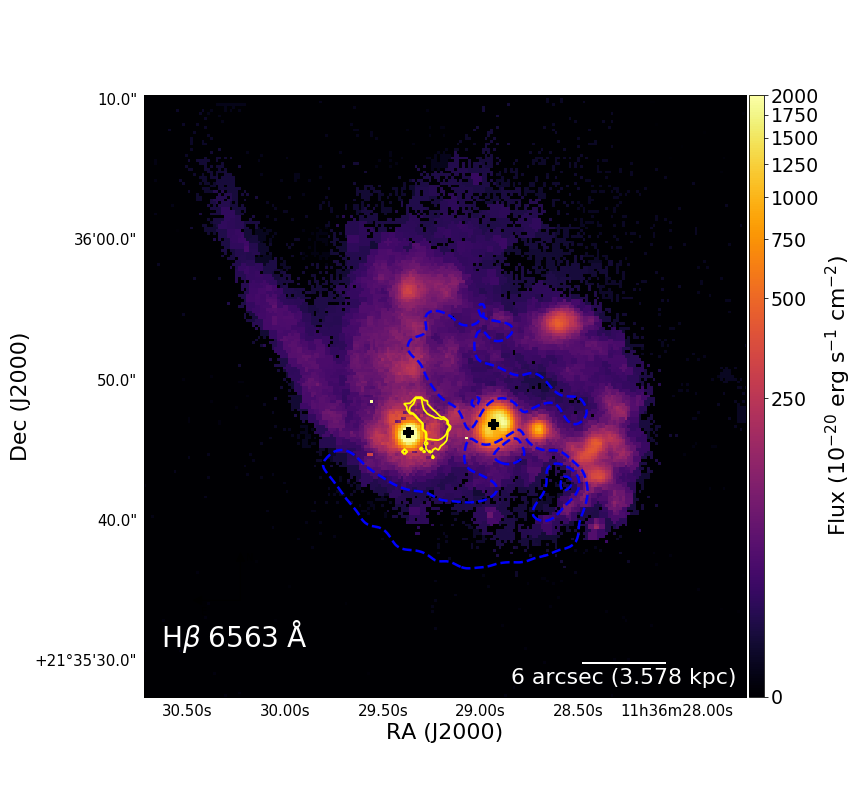}
\end{minipage}
\begin{minipage}[t]{0.49\linewidth}
\includegraphics[width=1.\linewidth,keepaspectratio=True]{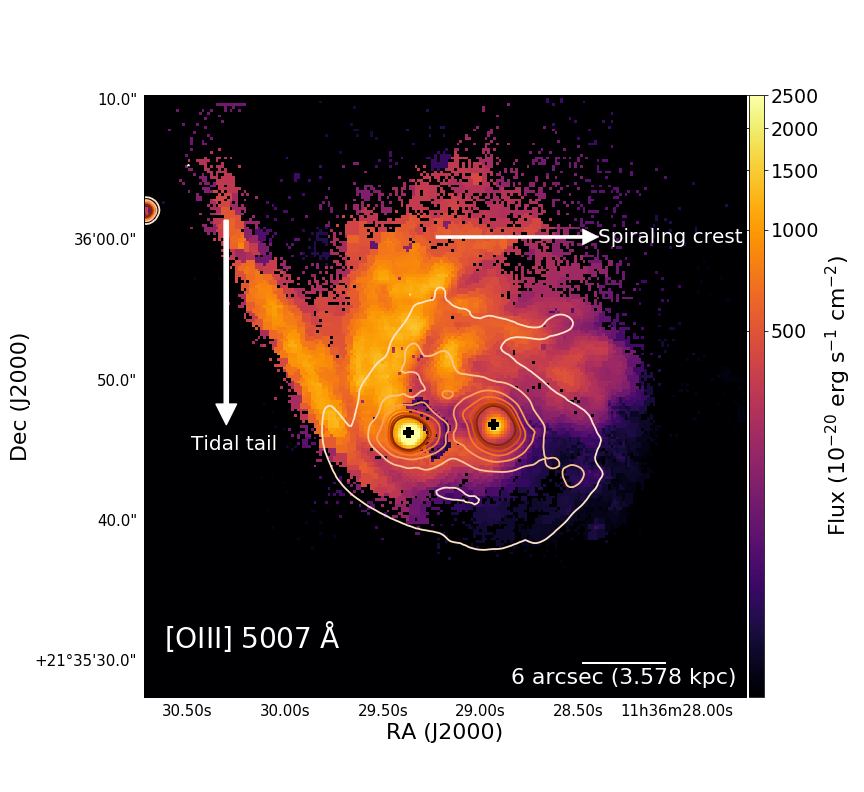}
\end{minipage}
\begin{minipage}[t]{0.49\linewidth}
\includegraphics[width=1.\linewidth,keepaspectratio=true]{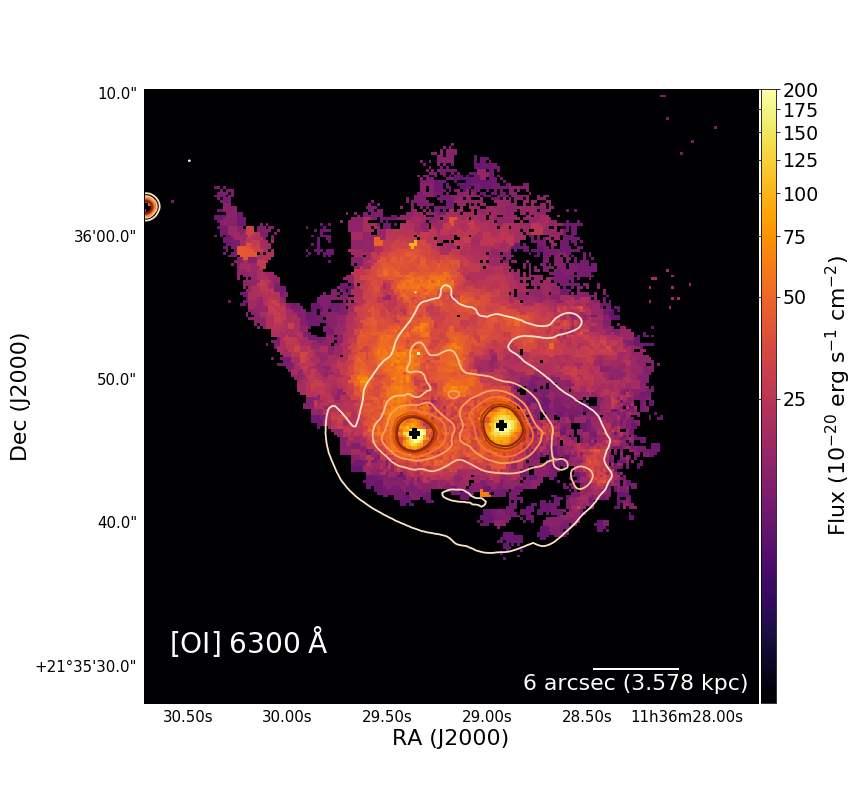}
\end{minipage}
\caption{Spatial distribution of the ionized gas in the dual AGN Mrk 739. \textit{Top panels:} Map of the flux of the narrow H$\alpha$  (\textit{left}) and  H$\beta$ (\textit{right}) atomic transitions from ionized gas. \textit{Lower panels:} Flux distribution maps for [OIII]$\lambda$5007 (\textit{left}) and [OI]$\lambda$6300 (\textit{right}). In all cases,  fluxes are given in units of $10^{-20}$erg s$^{-1}$ cm$^{-2}$. The line fluxes are computed by fitting single or double Gaussian components to all of the narrow lines. The scale bars, shown in the lower right corners of each panel, have an angular size of $6\arcsec$, corresponding to $\sim3.6$ kpc at the redshift of the source. The contours on the top-left and the lower panels represent the total optical emission, as shown in Figure \ref{fig:whiteimage}, while the dashed \textit{blue} contours in the top-right H$\beta$ emission map represent the regions with Balmer absorption features. The yellow contours in the H$\beta$ map highlight the region where double-peaked narrow-line emission is observed. In this region, we only report the values of the second component for all the emission maps. The black crosses mark the centers of the X-ray emission, as in Figure \ref{fig:whiteimage}. }
\label{fig:halphahbetaoiiioi}
\end{figure*}

\begin{figure*}[htp]
\begin{minipage}[t]{0.49\linewidth}
\includegraphics[width=1.\linewidth,keepaspectratio=true]{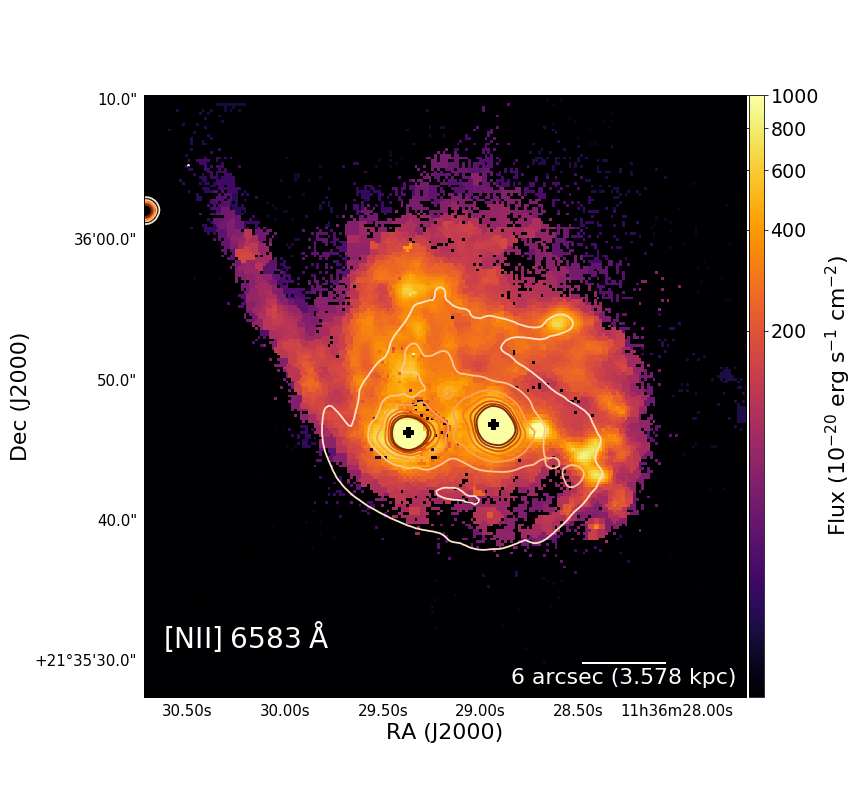}
\end{minipage}
\begin{minipage}[t]{0.49\linewidth}
\includegraphics[width=1.\linewidth,keepaspectratio=true]{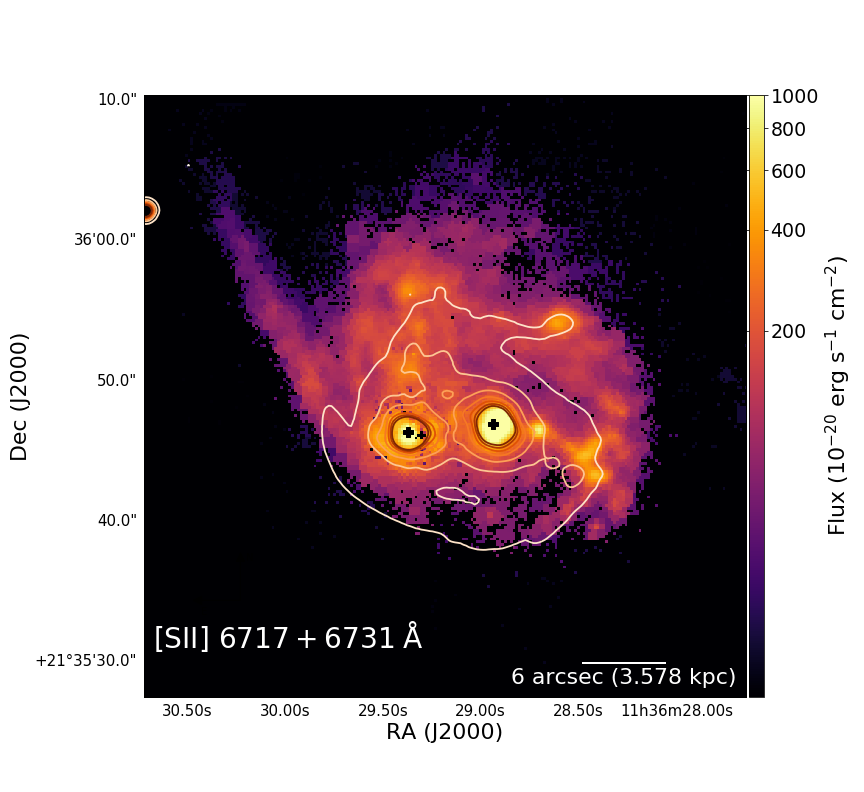}
\end{minipage}
\caption{Continuation of Figure \ref{fig:halphahbetaoiiioi}, with [NII]$\lambda6583$ ({\it left}) and the doublet [SII]$\lambda6717,\lambda6731$ ({\it right}). The scale-bar, contours, and symbols are the same as in Figure \ref{fig:halphahbetaoiiioi}.}
\label{fig:NIIbSII}
\end{figure*}

\subsection{Astrometric Calibration}\label{astrometrycorrection}

We astrometrically calibrated the VLT/MUSE data cube, using the sources reported by the Gaia Data Release 2 \citep{2018A&A...616A...2L} as reference. We matched the MUSE positions of Mrk 739E, Mrk 739W, and the star 2MASS J11363074+2136018 \citep{Monet_2003} that appears at the top-left corner of the field of view with the Gaia positions. The matching and offset determinations were carried out using the astrometrical calibration tool provided by the Aladin Sky Atlas \citep{2000A&AS..143...33B}. The measured offsets, corresponding to the average displacement of the three sources are $\Delta \rm RA=-1\farcs416\pm0\farcs042$ and $\Delta \rm Dec=0\farcs219\pm0\farcs016$. These offsets were applied in the subsequent analysis.

\section{Emission Line Analysis}\label{sec:results}

\subsection{Morphology of the Ionized Gas}\label{subsec:emission}

\begin{figure*}
\centering
\begin{minipage}[t]{0.4\textwidth}
\centering
\includegraphics[width=1\linewidth,keepaspectratio=true]{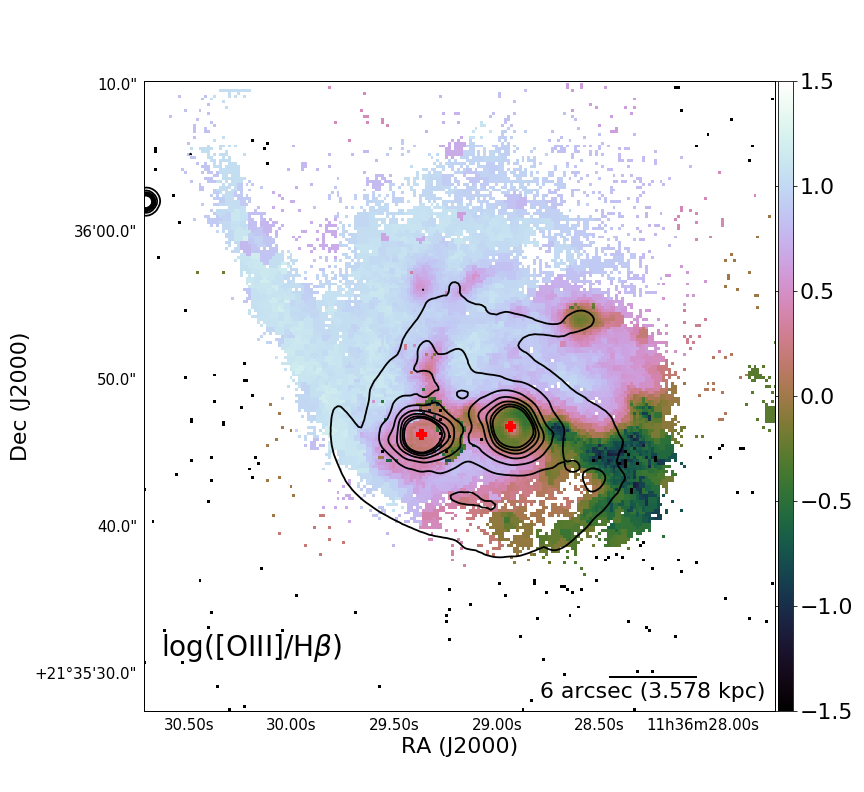}
\label{sii/halpha}
\end{minipage}
\begin{minipage}[t]{0.4\textwidth}
\centering
\includegraphics[width=1\linewidth,keepaspectratio=true]{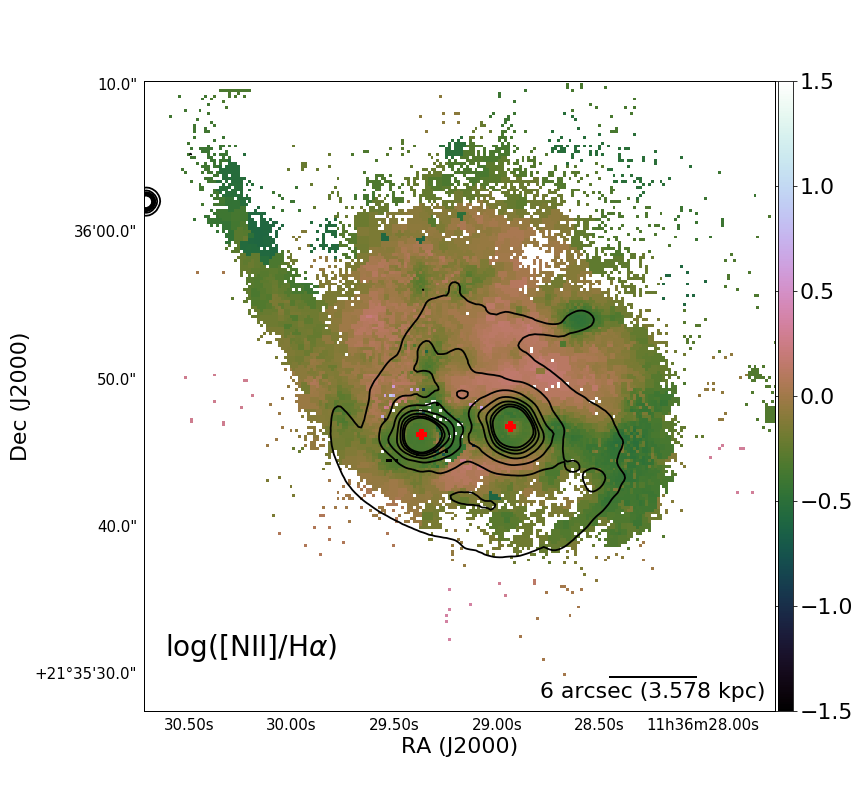}
\end{minipage}
\centering
\begin{minipage}[b]{0.4\textwidth}
\centering
\includegraphics[width=1\columnwidth]{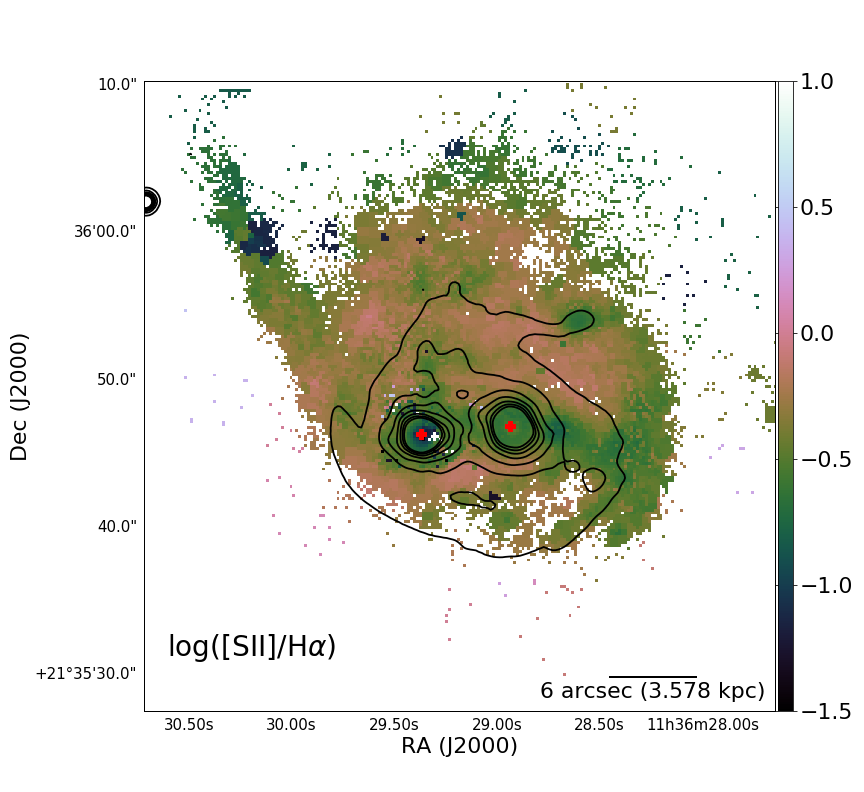}
\end{minipage}
\begin{minipage}[t]{0.4\textwidth}
\centering
\includegraphics[width=1\linewidth,keepaspectratio=true]{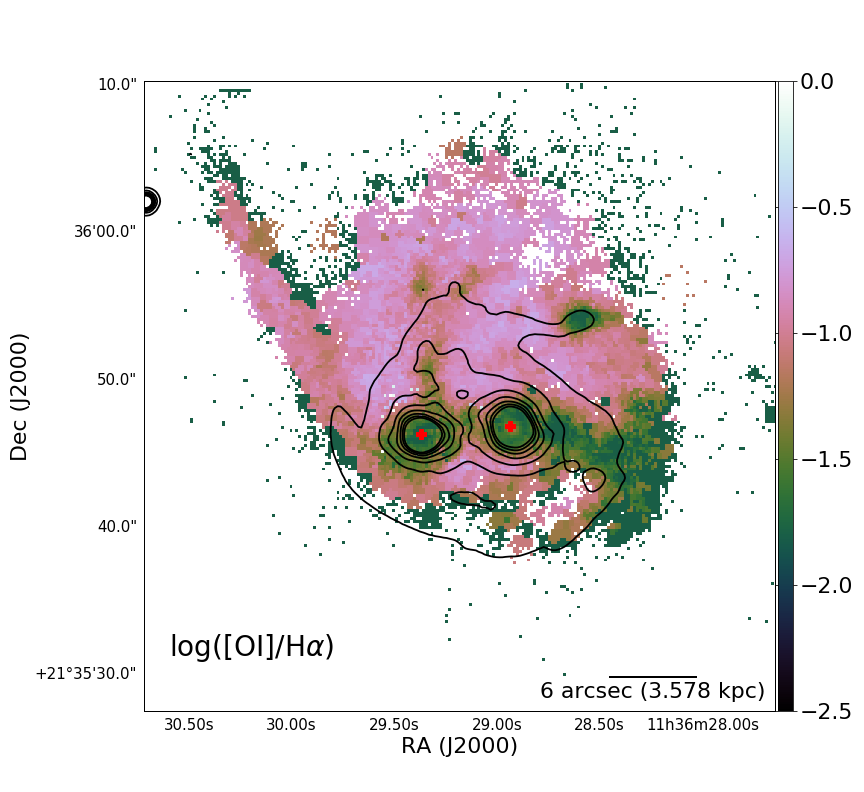}
\end{minipage}
\caption{Logarithmic emission line ratio maps for: [OIII]$\lambda5007$/H$\beta$ lines (\textit{upper-left}); [NII]$\lambda6583$/H$\alpha$ (\textit{upper-right}); [SII]$\lambda6717+\lambda6730$/H$\alpha$ (\textit{lower-left});  [OI]$\lambda 6300$/H$\alpha$ (\textit{lower-right}). The contours and symbols are the same as in Figure \ref{fig:halphahbetaoiiioi}. } 
\label{fig:emisions}
\end{figure*}

In order to understand the morphologies and kinematics for different atomic species, we analyze here the resulting maps for each emission line. Figures \ref{fig:halphahbetaoiiioi} and \ref{fig:NIIbSII} show the gas distribution for the main emission lines. We identify two clearly separated emission peaks, corresponding to the nuclei of each galaxy, and the arc-shaped clumpy structure observed in Balmer lines, [NII] and [SII], which is mostly located to the west of Mrk 739W. In these emission-line maps, especially in that of H$\alpha$ $\lambda6563$, we can see many features that are not present in the collapsed white-light image shown in Figure \ref{fig:whiteimage}, such as the fan-like stream of ionized gas in the north or several blobs associated with the aforementioned clumpy distribution to the west of the system. Below, in \S \ref{subsec:natureofionization}, we discuss the emission mechanism responsible for the emission of these regions using an optical emission-line diagnostic diagram in order to identify the origin of the ionized gas. The H$\beta$ emission is in general weak, particularly in the south region of the West nucleus, delimited by the \textit{cyan} contours, where the presence of H$\beta$ absorption lines overlaps with the emission line. The south of the East nucleus completely lacks H$\beta$ emission.

The spatial distribution of the [OIII]$\lambda5007$ (hereafter [OIII]) emission line starkly contrasts with the H$\alpha$ map discussed previously. The flux map clearly shows an extended and intense spiraling crest to the north of both nuclei, which does not have a symmetric equivalent to the south or west. Morphologically, neither [OIII] nor [OI]6300 reveal other obvious or distinguishing features. Other interacting pairs of galaxies show complex and disrupted features at more advanced stages like stellar bridges \citep{2020MNRAS.494.2785T}, irregular structures \citep{2017MNRAS.469.3629S}, or unique stellar halos \citep{2014MNRAS.442.3544F}, suggesting that the lack of those characteristics on Mrk 739 is evidence that we are seeing the merger either before or just after the first passage. The lack of [OIII] emission in the western region of Mrk 739E is due to we are only reporting the flux of the second component in this region. This choice is due to the relevance of the component either in the morphology, kinematics, and the evolutionary stage of the merger. The noteworthy extended tidal tail on the northeast of Mrk 739 is a ubiquitous feature of the gravitational interaction between galaxies that is formed by gas that is stripped from the outer regions of the interacting systems due to gravitational forces \citep{1972ApJ...178..623T,2008ApJ...683...94O,2015ApJ...807...73O}. In addition, Figure \ref{fig:NIIbSII} presents the distributions for the [NII]$\lambda6583$ (\textit{left}) line and the [SII]$\lambda6717,\lambda6730$ doublet (\textit{right}).

\subsection{Nature of the ionizing radiation}\label{subsec:natureofionization}

\begin{figure*}[htp]
\centering
\begin{minipage}[t]{1.0\textwidth}
\centering
\includegraphics[width=1\linewidth,keepaspectratio=true]{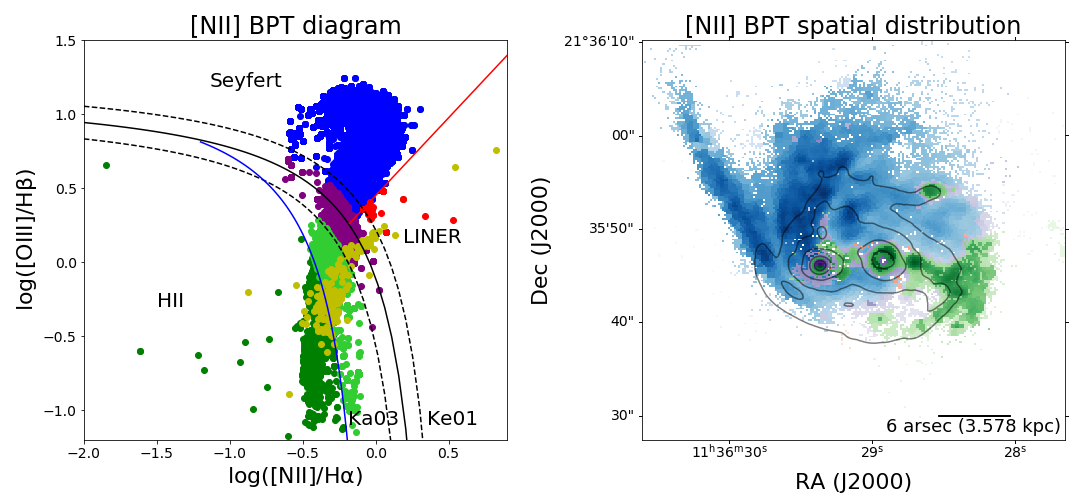}
\end{minipage}
\centering
\begin{minipage}[t]{0.49\textwidth}
\centering
\includegraphics[width=1\linewidth,keepaspectratio=true]{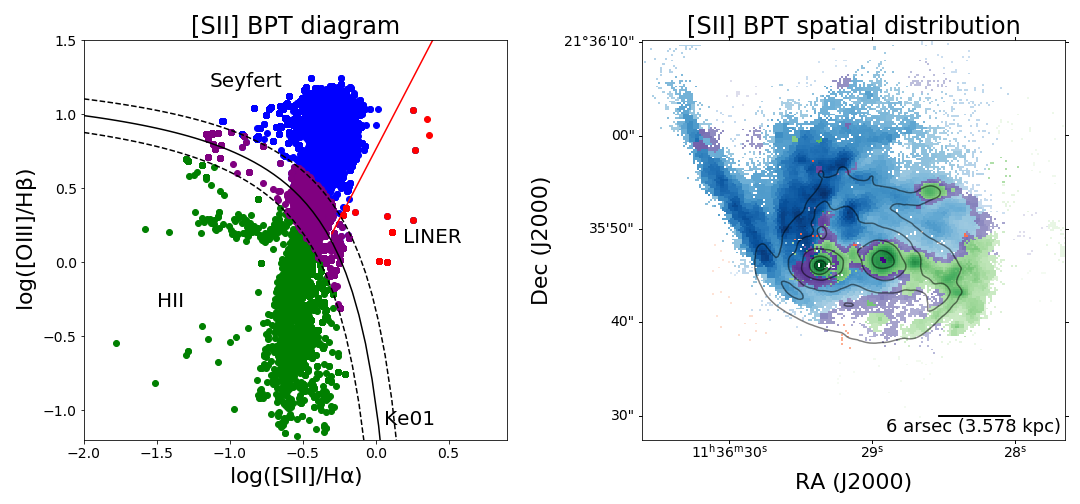}
\label{bptSii}
\end{minipage}
\begin{minipage}[t]{0.49\textwidth}
\centering
\includegraphics[width=1\linewidth,keepaspectratio=true]{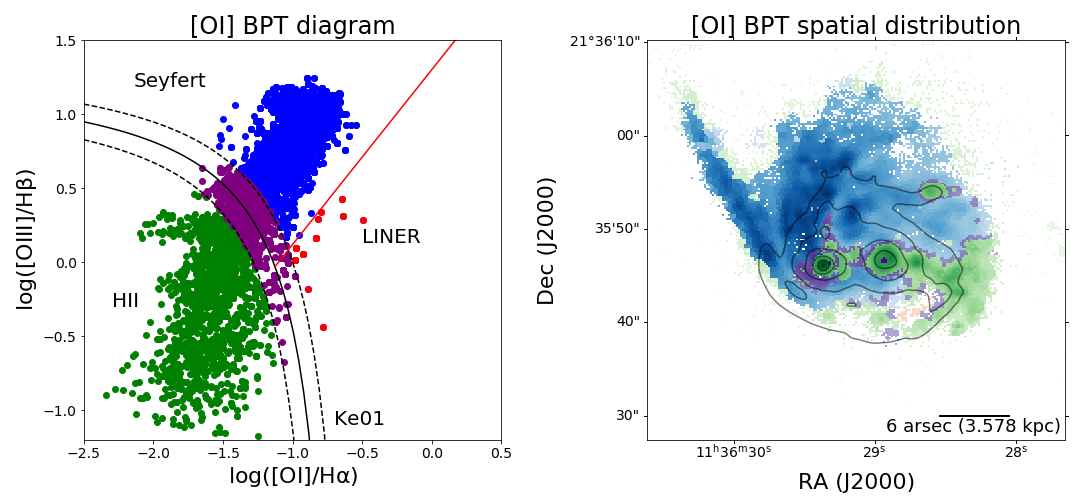}
\end{minipage}
\caption{BPT diagnostic diagrams and spatial distributions for Mrk 739. \textit{Top panels} show the [NII]-BPT ([OIII]$\lambda5007$/H$\beta$  versus [NII]$\lambda6583$/H$\alpha$) diagram (\textit{left}) together with the corresponding spatial distribution (\textit{right}). The \textit{lower panels} similarly present the [SII]-BPT ([OIII]$\lambda5007$/H$\beta$ versus [SII]$\lambda6717,\lambda6730$/H$\alpha$) and the [OI]-BPT ([OIII]$\lambda5007$/H$\beta$ versus [OI]$\lambda6300$/H$\alpha$) diagrams. The black solid curves mark the separation proposed by \citet[][denoted as Ke01 in the diagrams]{2001ApJ...556..121K}, between the theoretical maximum ionization driven by pure star-formation in HII regions and those regions ionized by AGN. Dashed curves represent $\pm 0.1$ dex of the division relation, corresponding to the estimated error for this threshold. The red solid line on the [NII]-BPT marks the separation between AGN and Low Ionization Narrow Emission-line Regions (LINERs), as proposed by \cite{2007MNRAS.382.1415S}, while the blue solid curve represents the separation between AGN and star formation reported by \cite{2003MNRAS.346.1055K} denoted as Ka03 on the [NII]-BPT diagram. The red solid line in [SII]-BPT and [OI]-BPT diagrams correspond to the separation between AGN and LINERS by \cite{2006MNRAS.372..961K}. In all panels, {\it blue circles} represent the spaxels where the AGN is expected to dominate the ionization, {\it green points} are the spaxels dominated by star formation processes and {\it red points} spaxels  in the LINERS locus.  {\it Purple symbols} correspond to those spaxels in the composite region, i.e., areas where the ionization can either be coming from AGN or star formation or a combination of them. {\it Lime Green dots} in the [NII]-BPT diagram denote the region between \cite{2001ApJ...556..121K} and  \cite{2003MNRAS.346.1055K}. \textit{Yellow dots} in the [NII]-BPT diagram correspond to those spaxels belonging to the double-peaked emission line region, which are not displayed on the resolved [NII]-BPT map. The resolved BPT diagrams are intensity-coded according to a particular line flux as follows: [OIII] for the blue regions ionized by AGN, H$\alpha$ for the green star-forming and purple intermediates regions, and [NII], [SII] and [OI] for the red LINERs regions.}
\label{fig:bpt}
\end{figure*}

We compute emission-line flux ratios using the best-fit narrow components for each individual spaxel with sufficient signal, as defined above. In Figure \ref{fig:emisions}, we present maps for the $\log$([OIII]$\lambda5007$/H$\beta$), $\log$([NII]$\lambda6583$/H$\alpha$), $\log$([SII]$\lambda(6717+6731)$/H$\alpha$), and $\log$([OI]$\lambda6300$/H$\alpha$) flux ratios, which we use to diagnose the nature of the ionization source \citep{1981PASP...93....5B,1987ApJS...63..295V,2000ApJ...530..704K,2001ApJ...556..121K,2006MNRAS.372..961K}. The green regions in the first two maps denote areas where H$\beta$ and H$\alpha$ dominate over [OIII] and [NII] lines respectively, which is indicative of star formation processes. On the other hand, the brown regions correspond to areas where the intensity is roughly equal for both lines. According to \cite{2006MNRAS.372..961K}, regions with $\log$([OIII]/H$\beta$)$>$1 are most likely dominated by AGN ionization. We identify a clear extended region that has high [OIII]/H$\beta$ ratio, being likely ionized by nuclear activity to the northeast.

We combine the optical emission-line ratios, discussed individually above, to obtain a unified classification of the dominating ionization source, as was originally done in the so-called Baldwin, Phillips \& Telervich diagram (BPT; \citealp{1981PASP...93....5B}). Figure \ref{fig:bpt} shows the BPT diagnostic diagrams for individual regions in the Mrk 739 system. The resolved BPT reveals an extended zone ionized by AGN activity and structures related to the western side of the galaxy merger, where star formation dominates. These maps reveal the widespread influence of the AGN, which is detected out to large distances ($5-20$ kpc) from the center. The innermost ($2-3$ kpc) surroundings of the two nuclei are consistent with being ionized by star formation, with composite ionization arising at the edges, where the transition between the central regions and the rest of the galaxy occurs. The yellow dots in the [NII]-BPT diagram correspond to the spaxels belonging to the double-peaked emission line, not shown on the maps for clarity. The region between \cite{2001ApJ...556..121K} and  \cite{2003MNRAS.346.1055K}, denoted by {\it lime green dots} in [NII]-BPT, characterize the edges of the most prominent star-forming regions in the resolved [NII]-BPT. There appears to be a nuclear region around Mrk 739E that is dominated by both composite and star formation mechanisms. This picture is consistent with the scenario proposed by \citet{2006ApJS..163....1H} where the gravitational interaction could have driven the gas toward the center, triggering star formation processes. In the particular case of Mrk 739E, we see that in the region of double-peaked emission line, both sets of lines are independently consistent with being ionized by star formation.

Finally, the clumpy western region is dominated by star formation, consistent with the results of \cite{2011ApJ...735L..42K}, who pointed out that Mrk 739W does not reveal evidence for an AGN in the optical/UV due to the high levels of star formation.

\subsection{Kinematics of the ionized gas}\label{subsec:kinematics}
\begin{figure*}
\begin{minipage}[t]{0.49\linewidth}
\includegraphics[width=1.\linewidth,keepaspectratio=True]{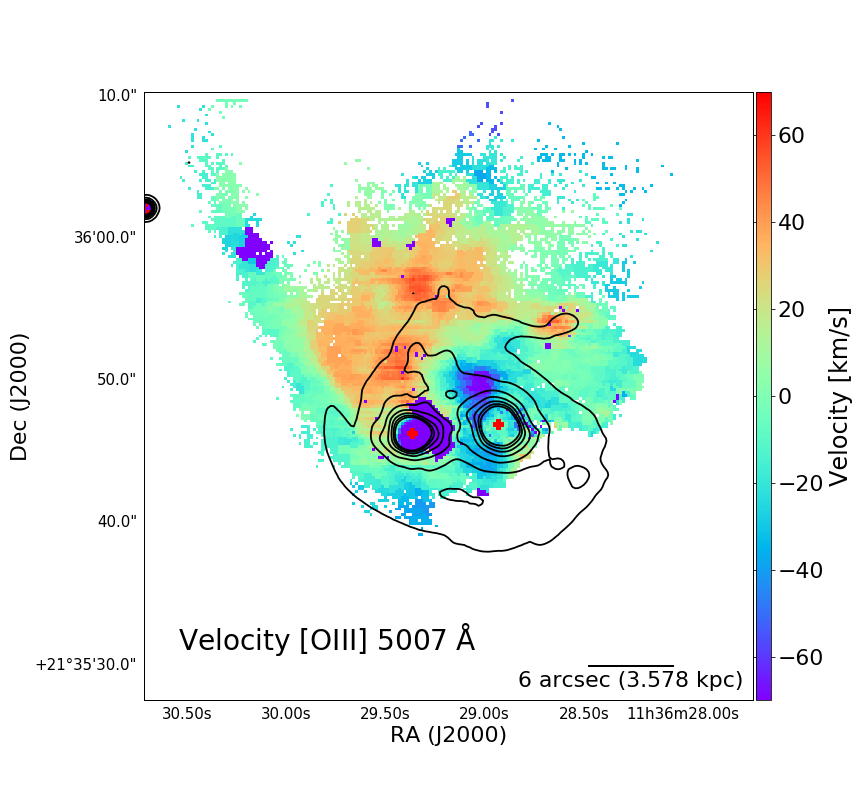}
\end{minipage}
\begin{minipage}[t]{0.49\linewidth}
\includegraphics[width=1.\linewidth,keepaspectratio=True]{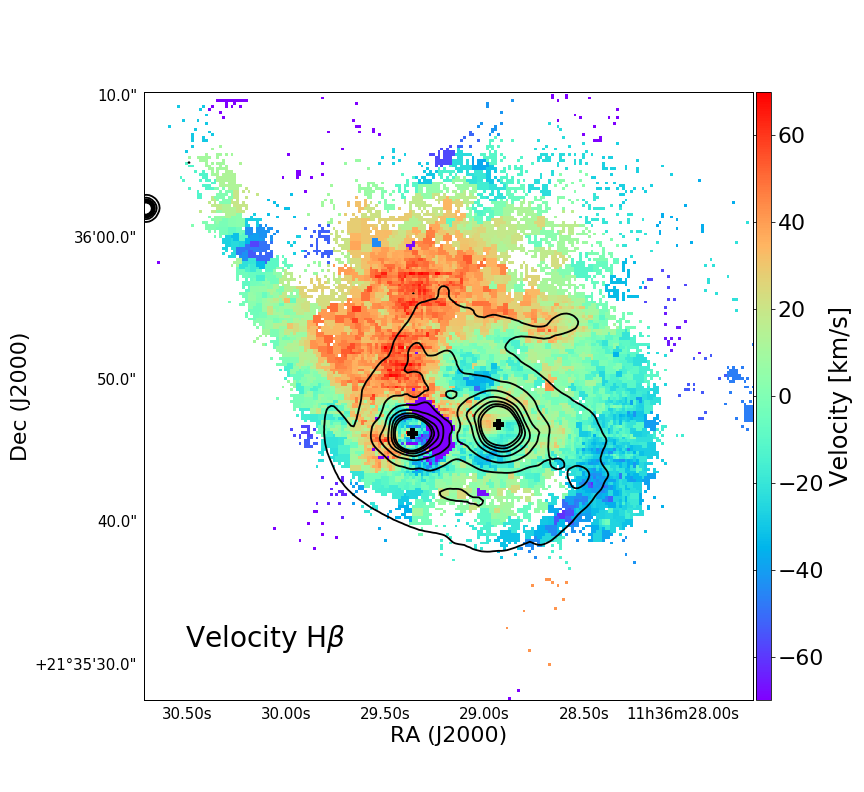}
\end{minipage}
\begin{minipage}[t]{0.49\linewidth}
\includegraphics[width=1.\linewidth,keepaspectratio=True]{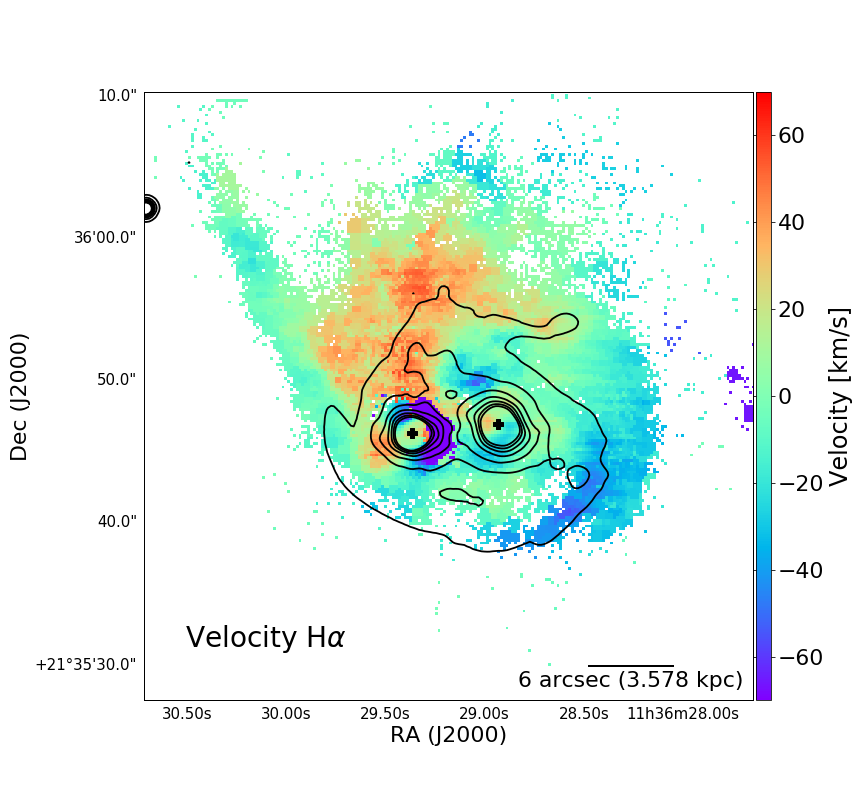}
\end{minipage}
\begin{minipage}[t]{0.49\linewidth}
\includegraphics[width=1.\linewidth,keepaspectratio=true]{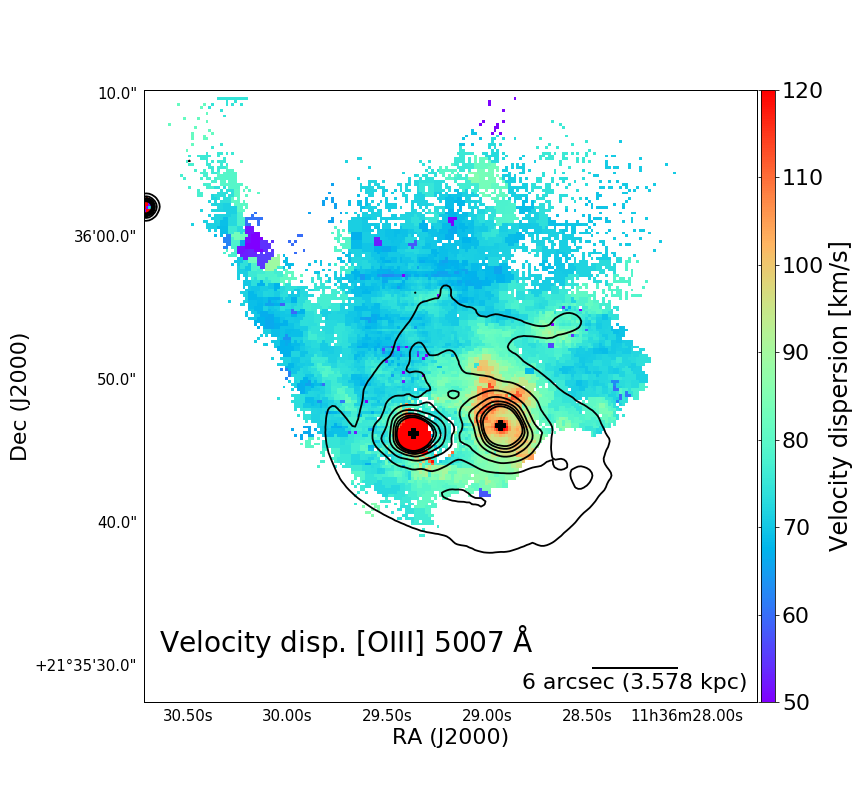}
\end{minipage}
\caption{Ionized gas kinematics in the dual AGN Mrk 739. Line-of-sight velocity maps are shown for [OIII]$\lambda5007$ (\textit{top-left}),  $\rm H\beta\;\lambda4861$ (\textit{top-right}) emission lines, and H$\alpha\;\lambda6563$ (\textit{lower-left}) emission lines, as well as the line-of-sight velocity dispersion map of the [OIII]$\lambda5007$ line (\textit{lower-right}). Velocities are presented in units of km s$^{-1}$. Black crosses mark the positions of the central nuclei of each galaxy, as measured in X-rays, while the contours correspond to the white-light emission, as shown in Figure \ref{fig:whiteimage}.}
\label{fig:kinematics}
\end{figure*}

The line-of-sight velocities and velocity dispersion of the gas were obtained from the Doppler shift and broadening of the three strongest emission lines: H$\beta$, [OIII]$\lambda5007$ and H$\alpha$. Their spatial distributions are presented in Figure \ref{fig:kinematics}, where we report the velocity values in the galaxy rest frame, which correspond to a heliocentric velocity of $8950\pm5$ km s$^{-1}$ based on $^{12}$CO($J=1\rightarrow0$) observations \citep{1996A&AS..116..193C}. As mentioned in \S \ref{sec:data}, the width of all emission lines was tied to that of [OIII]$\lambda5007$. For the [OIII], H$\beta$, and H$\alpha$ maps, we report the velocity obtained from the narrow Gaussian component that characterizes the systemic excited line emission of the galaxy and the velocity values of the second component, describing the aforementioned isolated structure that is ionized by star formation. Figure \ref{fig:kinematics} highlights a spatially extended region to the north, with velocities close to $\sim60$ km s$^{-1}$. Any motion associated with the eastern tidal tail, the western nucleus and the western side of the galaxy merger appears to be located in the plane of the galaxy, as all have negligible relative velocities. The nuclear region of Mrk 739E reveals a blue-shifted velocity of $\sim -130$ km s$^{-1}$ in [OIII] likely related to outflowing material coming from the eastern nucleus, resulting in a higher [OIII] velocity dispersion with values of $\sigma_{gas}\sim150$ km s$^{-1}$, as we see in the lower-right map of Figure \ref{fig:kinematics}. We see an intriguing blue-shifted region to the north of the western nucleus with values about $\sim -60$ to $\sim -85$ km s$^{-1}$ and a velocity dispersion between $\sim 100 -120$ km s$^{-1}$ extending north-south toward the Mrk739W nucleus. The velocity maps, in particular H$\alpha$, reveal a circular-like velocity profile surrounding the eastern nucleus with values ranging from $\sim -220$ km s$^{-1}$ to $\sim +40$ km s$^{-1}$ that might be related with a rotating disk. In the next subsection, we study in detail the origin of this component. 

\subsection{Double-peaked emission line region}\label{doubleline}

The MUSE data reveals a region around the east nucleus that appears to be partially decoupled from the rest of the system. The {\it top-right} panel of Figure \ref{fig:representativespectra} shows a clear double-peaked profile of the emission lines in that region. The presence of double-peaked emission lines in galaxies can be associated with galaxy mergers \citep{2018ApJ...867...66C} and dual AGNs \citep{Wang_2009}, or even outflows and rotating gas \citep{Greene_2005,2016ApJ...832...67N}. Taking advantage of our IFU data, in the \textit{lower-left panel} of Figure \ref{fig:kinematics} we morphologically identify a rotating disk-like velocity profile surrounding the East nucleus.

Spectrally, the double-peaked emission line region consists of a bluer emission line that becomes gradually redshifted as the position varies from east to west direction of the eastern nucleus. The originally redder line does not change its velocity with position. Instead, the flux of the redder line decreases until it disappears after reaching the eastern side of the disc, explaining thus the absence of double-peaked emission lines on the eastern side of the disk. We then consider the bluer moving line as the rotating component, while the redder and static line matches the velocity of the extended northern spiraling crest, and is thus related to the large-scale distribution of the galaxy. 

\begin{table*}[t]
\centering
\begin{tabular}{c c c c c} 
 \hline
 Parameter & Search range & Best fit & Error (68 per cent conf.) &Error (99 per cent conf.) \\ [0.5ex] 
 (1) & (2) & (3) & (4) & (5) \\ [0.5ex]
 \hline
  Scale factor (kpc)  & 0 $\rightarrow$ 4 & 1.21 & -0.01,+0.01&-0.03,+0.06 \\
 Flat Velocity (km s$^{-1}$) & 0 $\rightarrow$ 400 & 237 & -5.8,+ 5.9&-27.6,+25.5 \\
 Inclination ($^{\circ}$) & 0 $\rightarrow$ 89 & 33.10 & -0.93,+1.03 &-3.1,+4.9\\
 Position-angle ($^{\circ}$) & 90 $\rightarrow$ 180 & 156.99 & -0.39,+0.40&-1.64,+1.70 \\ [1ex] 
 \hline
\end{tabular}
\caption{Parameters and statistical uncertainties for the best-fit for the rotating disk, as obtained from the KinMS\_MCMC routine. Free parameters on our model are listed in Column 1. Columns 2 and 3 present the priors and the best fit of the posterior distribution with the errors of the best fit at 68 and 99 percent of confidence in Column 4, and 5.  }
\label{table:1} 
\end{table*}
There are two possible lines we might consider using to trace the kinematic structure around the eastern nucleus, [OIII] and H$\alpha$. The [OIII] emission line would be cleaner for the kinematic modeling, in order to avoid the broadening problem affecting the Balmer lines. However, the high extinction (see \S \ref{extinction}), the presence of blueshifted velocities in the nuclear region, and the spectral resolution of R$\sim1770$, corresponding to a velocity resolution (FWHM) of  $\sim170$ km s$^{-1}$ for the [OIII] line, are not sufficient to study the rotating disk map. The H$\alpha$ line, thanks to its higher S/N and spectral resolution of FWHM $\sim120$ km s$^{-1}$, allows instead for a better kinematic modelling, despite the presence of the BLR contaminating the central portion of the map.
  
In order to fit the velocity distribution of this rotating component around the East nucleus and derive its physical parameters, we perform a Markov Chain Monte Carlo (MCMC) kinematic simulation using the KINematic Molecular Simulation (KinMS\footnote{\url{https://github.com/TimothyADavis/KinMS}}) package developed by \citet{2013MNRAS.429..534D}. KinMS generates spectral cubes of synthetic data, used to simulate observations of arbitrary molecular/atomic gas distributions. We used it in order to create a simulated 1st-moment map, assuming a rotating disk, and then fit the associated parameters using the KinMSpy\_MCMC\footnote{\url{https://github.com/TimothyADavis/KinMSpy_MCMC}} routine.

We fit a relatively simple model, consisting of a rotating disk with an exponential surface brightness profile given by $\Sigma_{\rm H_{\alpha}}(r) = e^{\frac{r}{r_{0}}}$, where $r_{0}$ is the disk scale factor. We assume that the disk is rotating with a circular velocity profile that is constant with radius. This assumption is based on the fact that both spatial and spectral resolution do not allow to distinguish more complex features of the rotating disk. The line-of-sight inclination and position angle of the disk are free parameters of the model, along with the scale factor and the flat velocity of the rotating disk. The angles are measured starting from the north and moving counterclockwise.

\begin{figure*}
\centering
\includegraphics[width=0.8\textwidth,keepaspectratio=true]{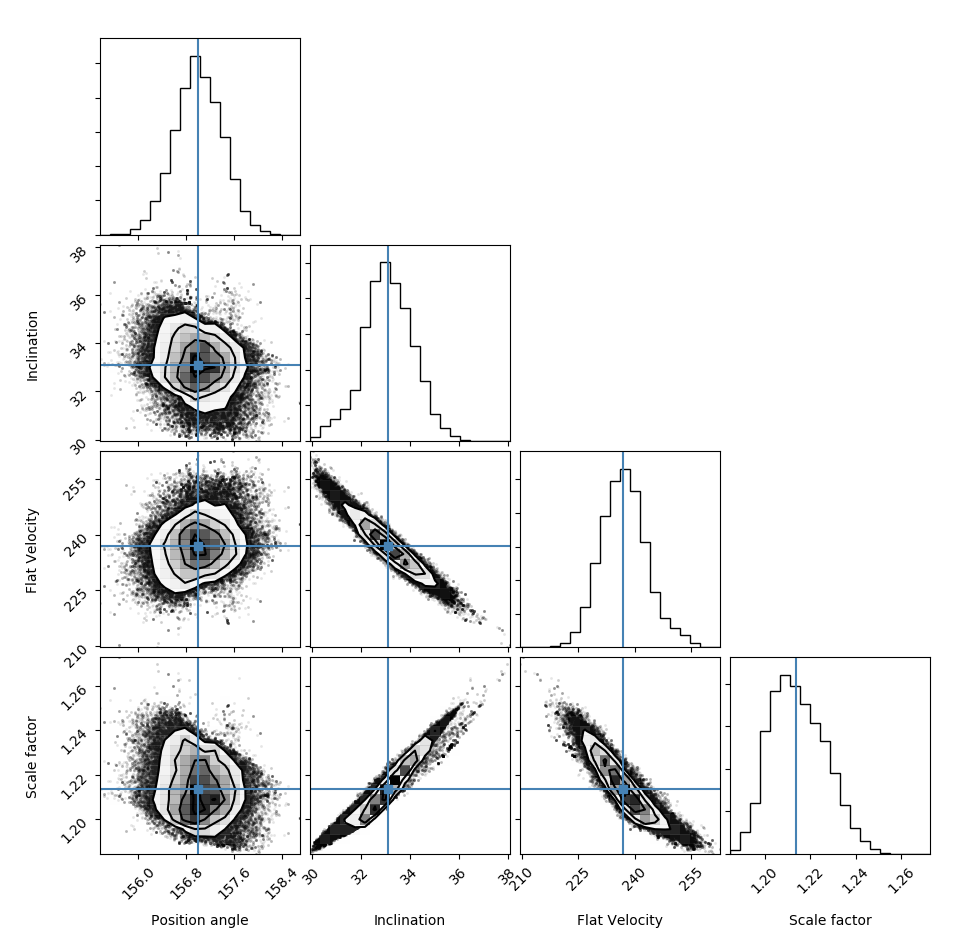}
\caption{Posterior distributions for each parameter fitted with the MCMC technique, as described in the text. Each panel shows the co-variation of a pair of parameters, while panels in the diagonal show the marginalized one-dimensional distribution for each parameter. The blue lines and squares mark the best-fit value for each parameter in our model. Degeneracies are clearly visible between the inclination and the flat velocity, the inclination and the scale factor, and the scale factor with the flat velocity. The position angle is the only fully independent parameter.}
\label{fig:mcmcimage}
\end{figure*}

\begin{figure*}
\begin{minipage}[t]{0.333\textwidth}
\includegraphics[width=1.0\linewidth,keepaspectratio=true]{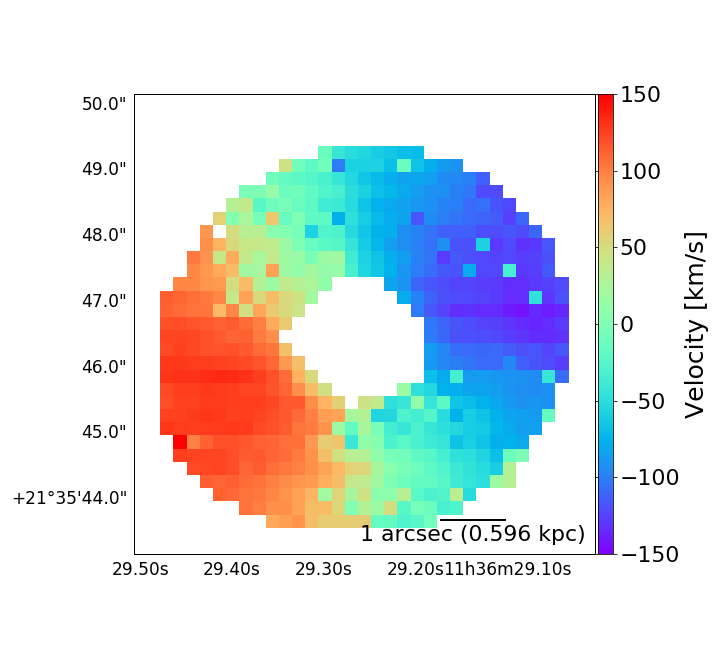}
\label{Halphavel}
\end{minipage}
\begin{minipage}[t]{0.333\textwidth}
\includegraphics[width=1.0\linewidth,keepaspectratio=true]{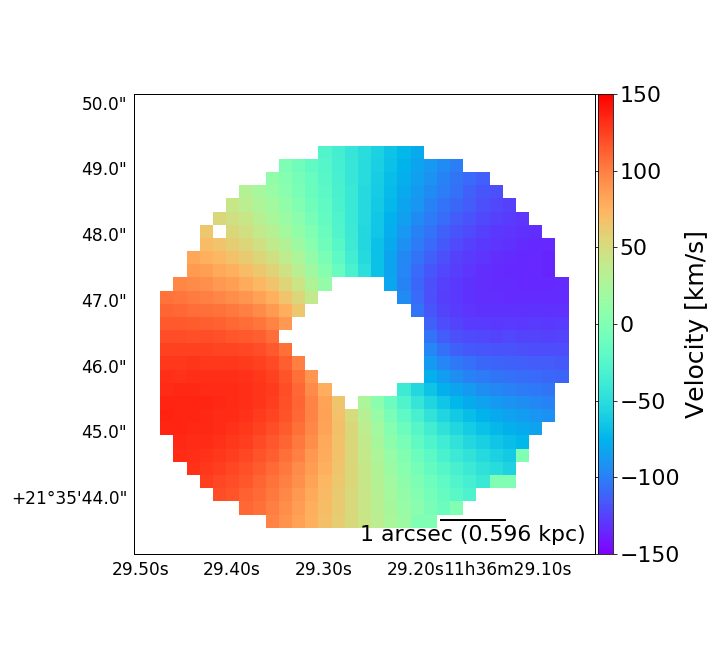}
\end{minipage}
\begin{minipage}[t]{0.333\textwidth}
\includegraphics[width=1.0\linewidth,keepaspectratio=true]{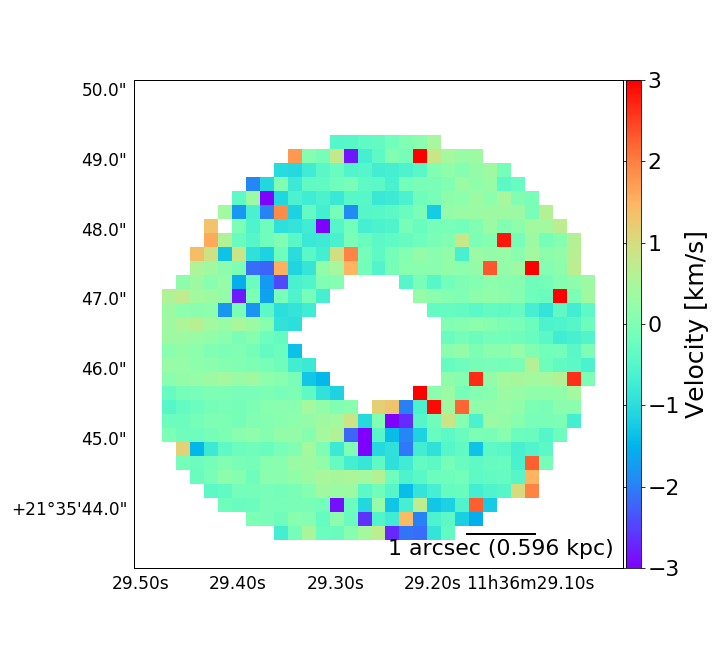}
\end{minipage}
\caption{\textit{Left:} H$\alpha$ velocity map of the rotating disk component found on the East nucleus, as presented in Figure \ref{fig:kinematics}. \textit{Center:} Best-fit kinematic model of the disk, obtained using the KinMS\_MCMC routine.  \textit{Right:} Residual maps of the fit calculated as the difference between the data and the model, divided by the RMS of the data.}
\label{fig:datamodelresiduals}
\end{figure*}

The Gibbs sampling MCMC framework with adaptive stepping implements a Bayesian analysis technique to explore the parameter space using a set of walkers from the emcee algorithm \citep{2013PASP..125..306F} to find the posterior distribution of the model parameters. A standard log-likelihood function, based on the chi-squared $\chi^{2}$ distribution $ln\;P = -\chi^{2}/2$, was used by each walker in order to maximize the likelihood and determine their next step through parameter space. Assuming Gaussian errors for our free parameters, we can estimate the goodness-of-fit of our model using $\chi^{2}$ statistics, $\chi^{2}=\sum_{i} (\frac{\rm data_{i}-\rm model_{i}}{\sigma_{i}})^{2}$ for each pixel, $i$. Since we are studying the 1st-moment of the velocity of the H$\alpha$ line profile, we need to compute the root-mean-square (RMS) noise, $\sigma$, in velocity units. To accomplish this, we made a new but simpler MCMC calculation over a few representative spaxels within the disk region, where the signal of the spectra was modified by random Gaussian noise. The variation was spectrally convolved with a Gaussian kernel matching the spectral resolution of the MUSE instrument. We then fitted the convolved spectra with Gaussian components, as described in Section \ref{sec:data}. We repeated this process one thousand times to obtain statistically significant results. For each spectrum, we calculated the line of sight (LOS) velocity for all relevant emission lines. We then took the standard deviation of the thousand velocities as noise. We found that the noise associated with the measurement of the velocity is mostly constant, at a value of $\sim$20 km s$^{-1}$. In order to exclude from the fit the inner region, dominated by the emission from the BLR associated with the AGN, we assign here artificially high RMS values.

Our MCMC fitting procedure considers $1.5\times10^{5}$ steps, using uniform priors in linear space. The priors, guesses, and resulting best values for the parameters of our model are presented in Table \ref{table:1}. In addition, Figure \ref{fig:mcmcimage} shows the so-called corner plot, where we can see the posterior distributions of parameters together with the one-dimensional marginalization of the physical parameters that characterize the rotating disk. There is a clear degeneracy between the inclination of the disk and the circular velocity since the relation between the two is given by $vel\propto\frac{vel_{obs}}{sin (i)}$. Also, the scale factor is degenerate with the inclination, due to projection effects on the plane of the sky, while the scale factor correlates with the final circular velocity because of the dependency of the velocity with the inclination.

For easier visual inspection, in Figure \ref{fig:datamodelresiduals} we present the kinematically-detached rotating disk of Mrk 739E along with the best-fit model and their residuals. We find a well-defined rotating disk, with a circular velocity of $237^{+26}_{-28}$ km s$^{-1}$ inclined by $33^{+5}_{-3}$ degrees. The model agrees reasonably well with our data, considering the small velocity residuals presented in the right panel of Figure \ref{fig:datamodelresiduals}. We expect that these low residual values, mainly in regions where the disk velocity is close to 0 km s$^{-1}$, are produced by a non-optimal separation in our spectral fitting procedure of the disk component and the component related to the galaxy, due to the spectral resolution. The proper convergence of the parameters within our MCMC treatment is not affected by those residual values.  Multi-wavelength IFU data at higher spatial resolutions, as those obtained using the MUSE Narrow Field Mode, as part of ESO program 0104.B-0497(A), would be helpful to understand and better constrain the physical parameters of the rotating disk and disentangle the central AGN-dominated emission from the disk. 

Our simple rotating disk model allows for the determination of the mass of the material surrounding the East nucleus just considering Newtonian physics. Using the flat velocity and scale factor $r_{0}$ as a characteristic radius, we can find that the mass within $r_{0}$ is given by:

\begin{equation}
    M(\rm R)= v^{2}\frac{r_{0}}{G}
\end{equation}

\noindent 
where $G$ is the gravitational constant, $v$ and $ r_{0}$ are the flat velocity and the scale factor summarized in Table \ref{table:1}. Thus, inside a radius of $1.21$ kpc centered at the East AGN position, the mass is $\log M(M_{\odot})=10.20\pm0.06$, almost three orders of magnitude larger than the SMBH mass we estimate for the East nucleus in $\S \ref{smbhmass}$ based on our single epoch BLR constraints.

Unlike other dual AGN, e.g., NGC 6240 \citep{M_ller_S_nchez_2018,2020A&A...633A..79K}, Mrk 463 \citep{2018ApJ...854...83T}, and Mrk 266 \citep{2012AJ....144..125M}, Mrk 739 does not show evidence of large-scale structures with high velocity dispersion generated by the ongoing major merger. The system presents narrow lines with a velocity dispersion close to the spectral resolution of MUSE at large-scales, while the nuclear region of Mrk 739E displays high velocity dispersion in [OIII], likely related to an outflow with a dispersion of $\sigma_{\rm gas}\sim 170$ km s$^{-1}$. The dispersion in Mrk 739W is more extended with values close to $\sigma_{\rm gas}\sim 120$ km s$^{-1}$. The rotating disk presents a uniform circular velocity and an inclination that indicates that this structure is kinematically decoupled from the rest of the merger.

\section{Stellar populations}\label{stellarpopulation}

The MUSE observations also allow us to trace the characteristics of the stellar populations within Mrk 739, based on the measurements of the optical continuum and absorption features. In order to proceed with such study, and as it is mentioned in \S\ref{sec:data}, we first made a Voronoi tessellation, with a target SNR of 40 per resolution element (1.25 \AA) over the full observe-frame spectral window since we are concerned with the fitting of the full spectrum. The binned spectra were analyzed using pPXF \citep{Cappellari2017} to derive parameters such as Line-Of-Sight velocity, velocity dispersion, stellar age, and stellar metallicity from the stellar continuum. Our procedure uses templates of single stellar populations (SSP) from the extended MILES (E-MILES) library, which cover the full spectral range between  1680-50,000 \AA~at moderately high-resolution \citep{2016MNRAS.463.3409V}. In the particular case of Mrk 739, we employed scaled-solar theoretical Padova00 isochrones \citep{2000A&AS..141..371G}, a unimodal IMF with a slope of 1.3, and solar $\alpha$-elements abundances from the E-MILES templates (see, e.g., \citealp{2016MNRAS.463.3409V} for references).

A bootstrap approach was followed in order to statistically retrieve the uncertainties on the measurements of the kinematics and the stellar populations for each Voronoi-binned spectrum. The routine produces 1120 Monte Carlo simulations for each binned spectrum by adding random noise and fitting with pPXF. In order to do this, a first-pass spectral-pixel mask is initially obtained from the original spectrum, by conservatively masking the typical gaseous emission lines in the fitted rest-frame wavelength window (i.e., 4650-8980 \AA). Additionally, three spectral regions are masked to prevent contamination by AGN broad emission lines and poorly corrected telluric lines: 4680-4690 \AA, 5870-5885 \AA, and 7350-7450 \AA. The pixel mask is then complemented using a sigma-clipping at a 5$\sigma$-level, after one preliminary pPXF fit. To mask sky subtraction residuals affecting the red end of each spectrum, an additional sigma-clipping is performed by making use of a more stringent threshold, which has been selected mostly to allow pPXF to fit the CaII triplet absorption lines. 

\begin{figure*}[htp]
\centering
\includegraphics[width=1.0\textwidth,keepaspectratio=true]{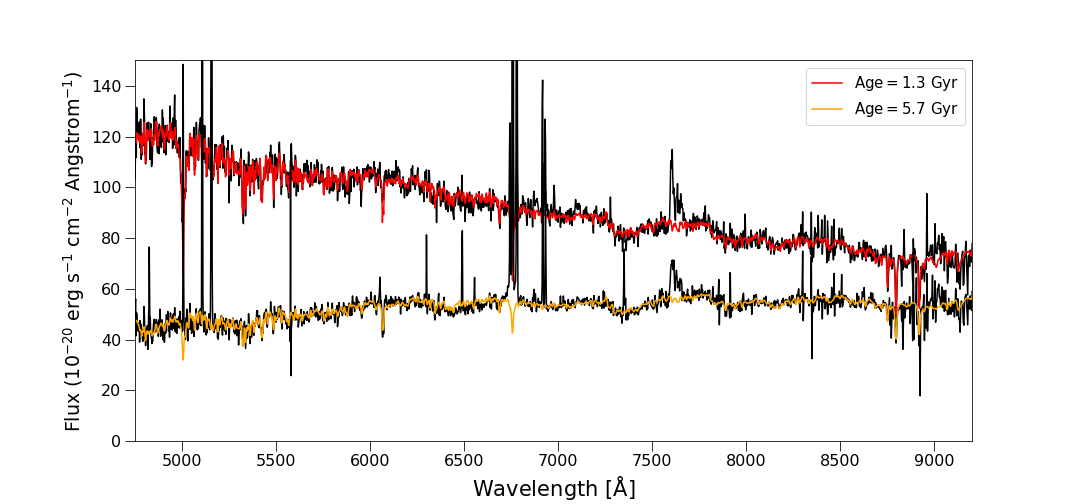}
\caption{ Representative stellar population fits from the extracted regions shown in {\it top-left panel} of Figure \ref{fig:SPsn40padova}. The stellar continua of the observed spectra ({\it black curve}) are fitted with the pPXF software, as described in \S \ref{stellarpopulation}. The best fit for the top spectrum is characterized by a young stellar population with an age of $\sim$1 Gyr ({\it red curve}), while for the bottom spectrum we obtain an old stellar content, with an age of $\sim$6 Gyr ({\it orange curve}).}\label{fig:specrasp}
\end{figure*}

\begin{figure*}[hb]
\begin{minipage}[t]{0.5\textwidth}
\includegraphics[width=1.0\linewidth,keepaspectratio=true]{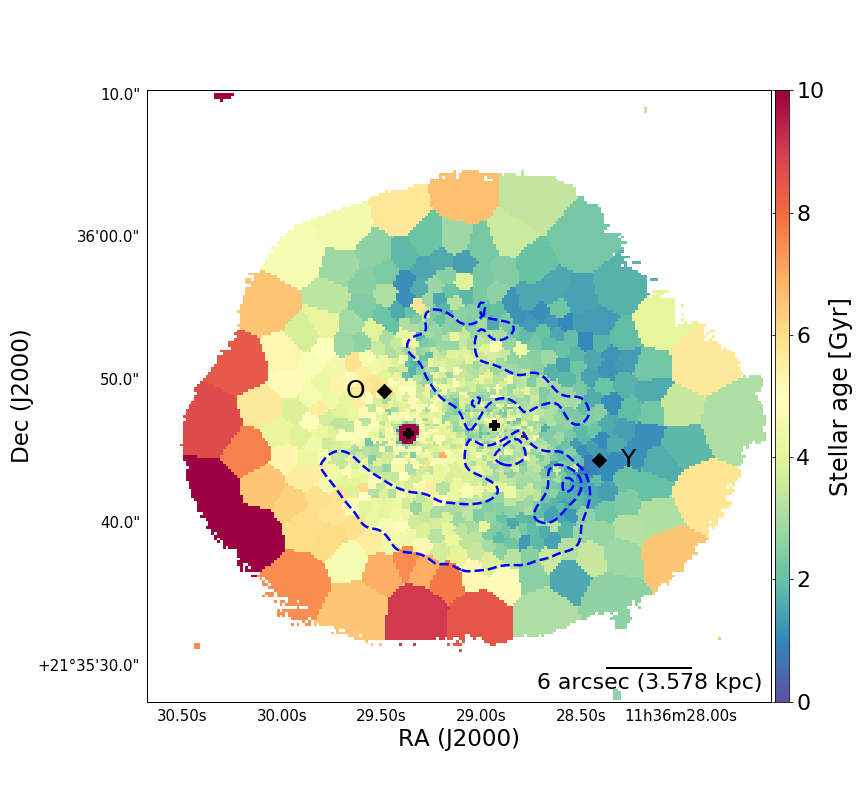}
\end{minipage}
\begin{minipage}[t]{0.5\textwidth}
\includegraphics[width=1.0\linewidth,keepaspectratio=true]{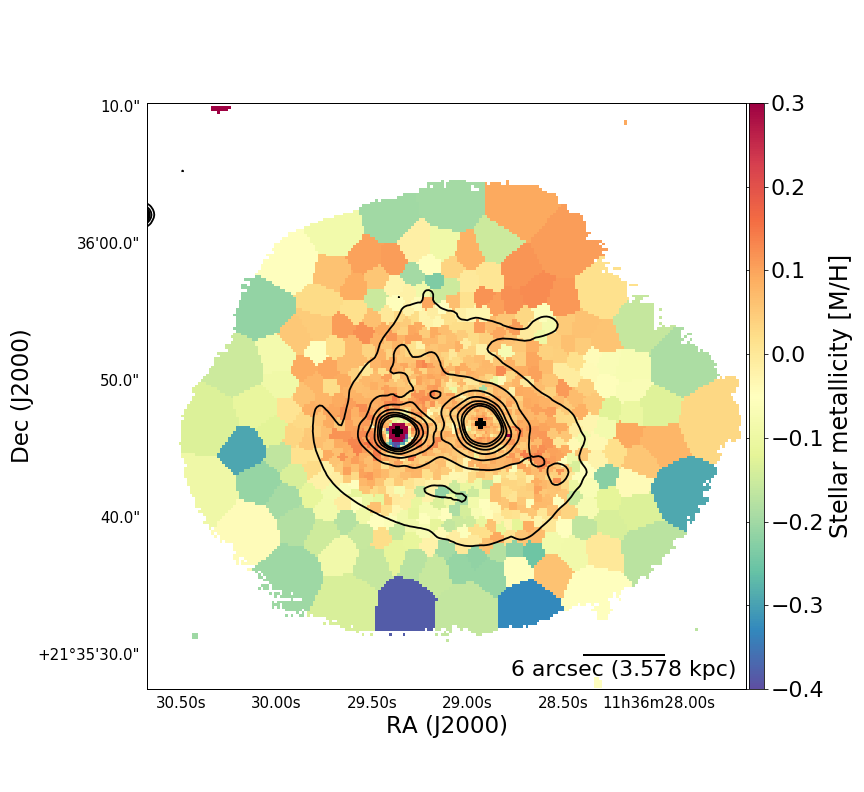}
\end{minipage}
\begin{minipage}[t]{0.51\textwidth}
\includegraphics[width=1.0\linewidth,keepaspectratio=true]{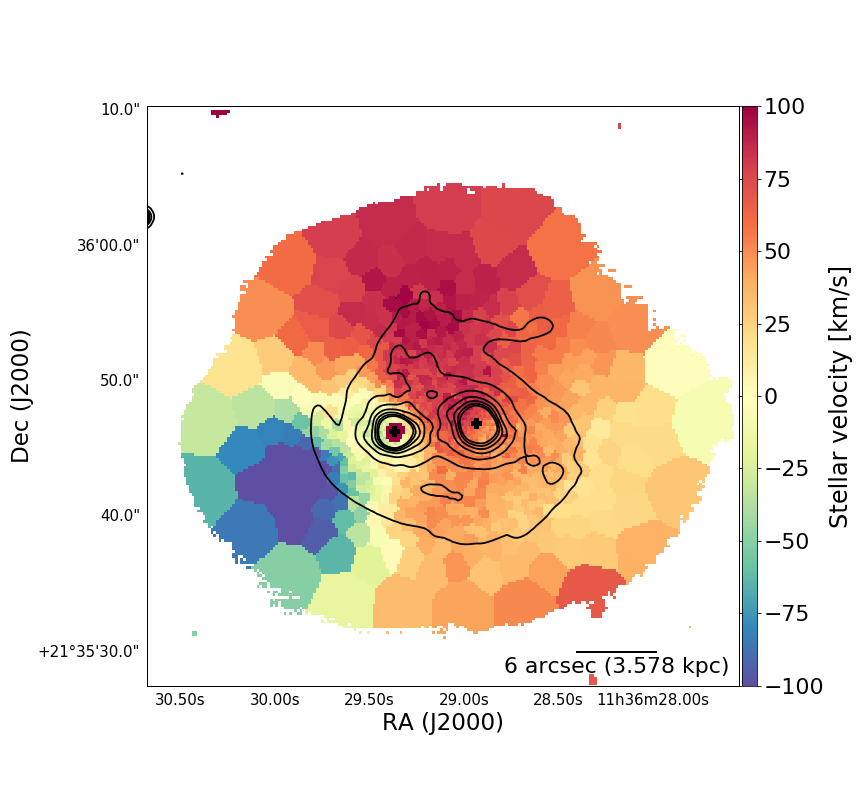}
\end{minipage}
\begin{minipage}[t]{0.5\textwidth}
\includegraphics[width=1.0\linewidth,keepaspectratio=true]{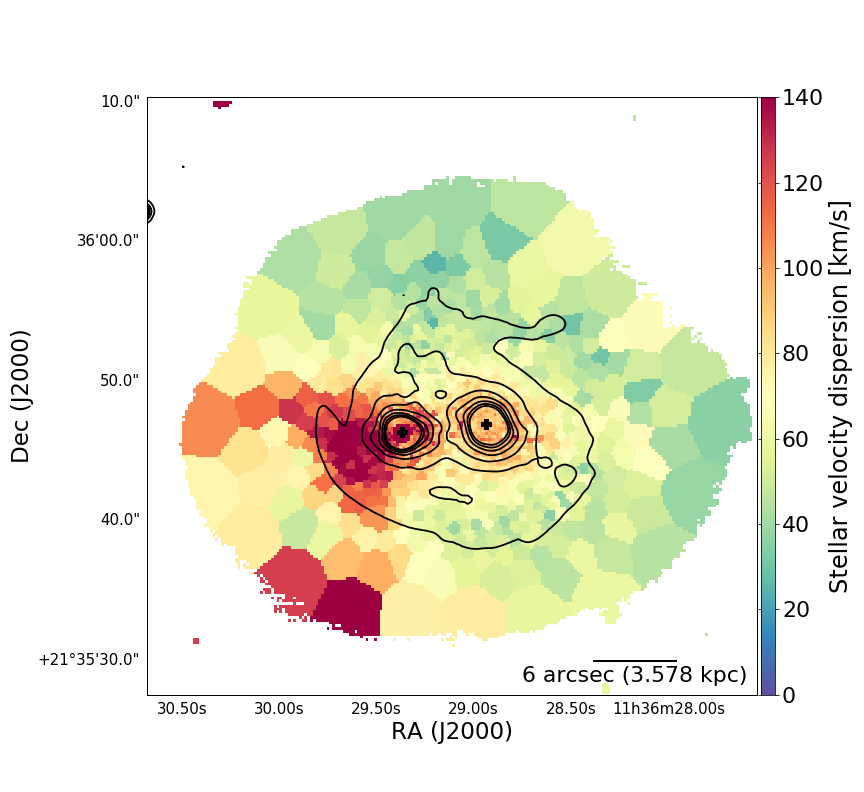}
\end{minipage}
\caption{Map of the stellar population parameters,  as obtained using our full spectral fitting routine. \textit{Top panels:} Stellar ages in Gyr ({\it left}) and metallicity ({\it right}). The metallicity ranges between -0.3 and +0.33, relative to the solar metallicity. \textit{Lower panels}: Stellar velocity ({\it left}) and velocity dispersion ({\it right}). The kinematics are measured in km s$^{-1}$, while the LOS velocities are reported relative to the systemic velocity of the system. The \textit{blue} contours on the \textit{upper-left} panel highlight the regions where we found Balmer absorption features,  while the diamond symbols mark the position of the spectra presented in Figure \ref{fig:specrasp}. The region marked with ``Y'' corresponds to the young stellar population, top spectrum in Fig.~\ref{fig:specrasp}, while the ``O'' corresponds to the old stellar population, bottom spectrum in Fig.~\ref{fig:specrasp}. The black crosses mark the centers of the X-ray emission, as in Figure \ref{fig:whiteimage}.}\label{fig:SPsn40padova}
\end{figure*}

\begin{figure*}
\begin{minipage}[t]{0.5\textwidth}
\includegraphics[width=1.0\linewidth,keepaspectratio=true]{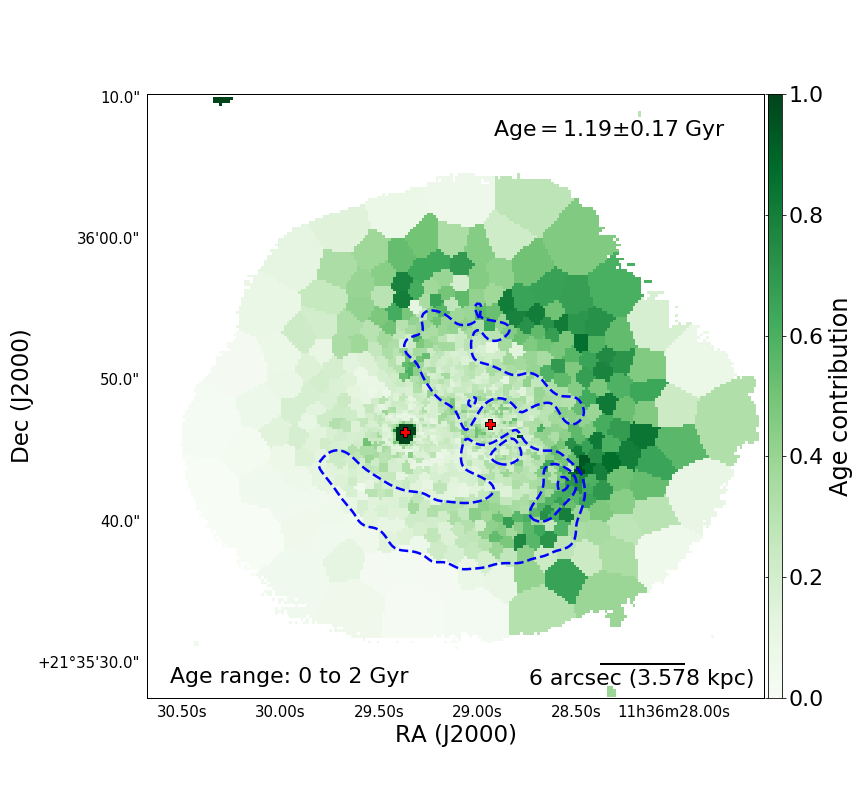}
\end{minipage}
\begin{minipage}[t]{0.5\textwidth}
\includegraphics[width=1.0\linewidth,keepaspectratio=true]{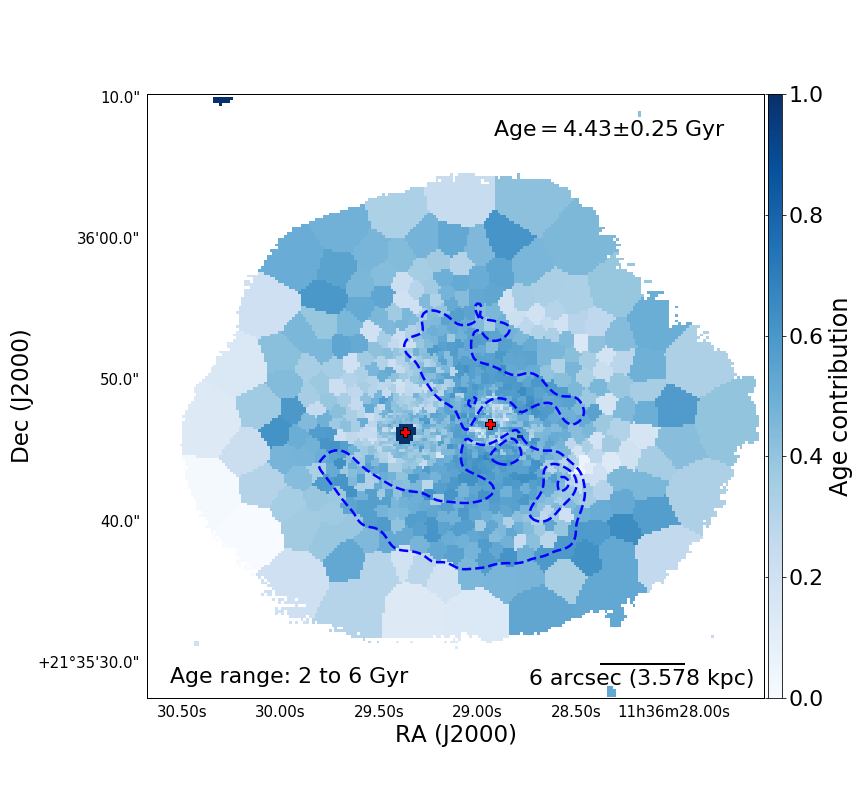}
\end{minipage}
\begin{minipage}[t]{0.51\textwidth}
\includegraphics[width=1.0\linewidth,keepaspectratio=true]{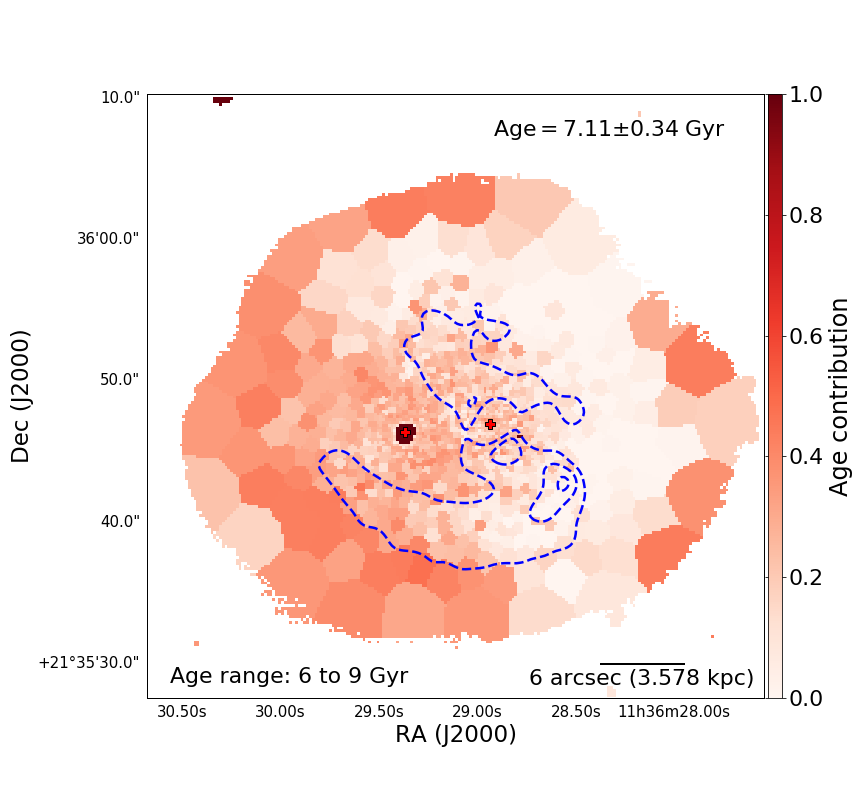}
\end{minipage}
\begin{minipage}[t]{0.5\textwidth}
\includegraphics[width=1.0\linewidth,keepaspectratio=true]{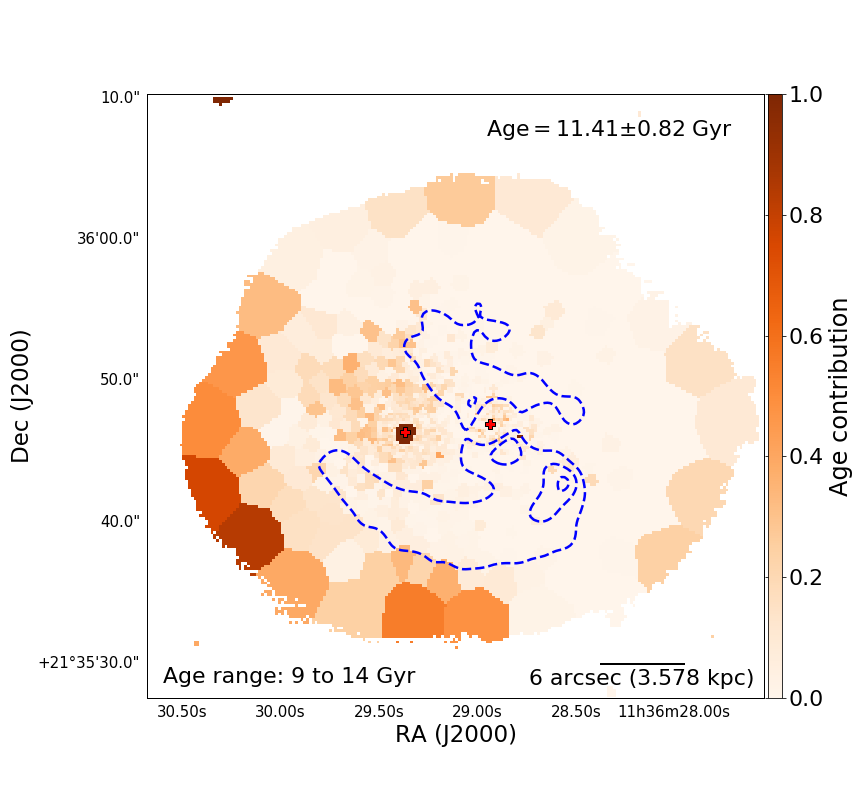}
\end{minipage}
\caption{Maps of the fractional contribution of the corresponding stellar populations to the best-fit template, in the following age bins: $0-2$({\it top-left}) , $2-6$ ({\it top-right}), $6-9$ ({\it lower-left}), and $9-14$ Gyr ({\it lower-right}). The \textit{blue} contours highlight the regions where we found Balmer absorption features. The median age and corresponding standard deviation for each map are shown in the {\it upper-right} corner.}\label{fig:agecontribution}
\end{figure*}

Then we adopted two different procedures in order to retrieve kinematics and stellar population parameters. Only additive polynomial and stellar templates were used to retrieve the stellar kinematics to avoid template mismatch. For the stellar population, we used stellar templates along with a multiplicative polynomial, which, unlike the additive polynomial, do not affect the line strength of the spectral features \citep[see, e.g.,][]{Cappellari2017}. To choose optimal additive and multiplicative polynomial degrees, we ran independent tests for the kinematics and the stellar population procedures, respectively. We let the polynomial degrees vary from 2 to 50 and achieved stable results for an additive polynomial degree of 16 (kinematics) and a multiplicative polynomial degree of 12 (stellar population). We adopted such degree values for the rest of the analysis. We fit the Voronoi-binned spectra by masking them according to the pixel masks retrieved in the aforementioned first stage and by making use of the optimal polynomial degrees.

In order to retrieve uncertainties, we implemented two different bootstraps following the approach of using exclusively additive or multiplicative polynomials for the kinematics and stellar populations, respectively. During the bootstrap, we randomized the choice of the parameters around the optimal values for each of the Monte Carlo realizations of the Voronoi-binned spectra. The randomized parameters are the additive polynomial degree (between 14 and 18), multiplicative polynomial degree (between 10 and 14), kinematic initial guess (around the previously measured values of velocity and velocity dispersion), and regularization parameter (between 90 and 100). The random choice of fitting parameters helps to avoid a possible bias introduced by the routine. Since we are dealing with spectra affected by strong emission lines  --- especially in the regions close to the two AGNs --- we only consider those spectra for which less than half of the spectral elements have been masked. Otherwise, we discard the corresponding Voronoi-binned spaxels as we cannot retrieve reliable measurements from that region. We also discarded those spectra where the fits are unconstrained by large errors.

Figure \ref{fig:specrasp} shows two representative spectra and their corresponding stellar population fits, for a young, $\sim$1 Gyr ({\it red curve}) and an old, $\sim$6 Gyr ({\it orange curve}), stellar population present in regions of the dual AGN Mrk 739. The best-fits agree very well with the observed data, proving the effectiveness of our procedure. We note that the Balmer absorption is stronger in the young stellar population, since it is produced by relatively young stars with ages $\sim 1$ Gyr \citep{2018MNRAS.477.1708P}.  

The kinematic maps and physical parameters of the stellar populations are shown in Figure \ref{fig:SPsn40padova}. The age and metallicity values are derived by taking the mean of the age and metallicity distribution of the best-fit templates, light-weighted by their contribution to the fit. Mrk 739 presents a population with ages ranging from $\sim$ 1 to 6 Gyrs and metallicities of [M/H] $\sim$ 0.1 in the central regions of the merger. The youngest stellar populations are located on the western side of the galaxy, coincident with the most extended star-forming region, as characterized in \S \ref{subsec:natureofionization}. Older stars reside on the eastern side of the galaxy, where the stellar age reaches values up to 5 - 6 Gyrs. The oldest ($\sim 7-10$ Gyr) population at the southeast edge has lower metallicities ([M/H] $\sim -0.25$) and distinctive kinematics dominated by high velocity dispersion and blueshifted velocities. The blue contours in the stellar age map highlight regions where the Balmer absorption feature is strong, indicating a dominant population with ages between $\sim$ 1.4 to 4.3 Gyr. The southern portion of the contoured Balmer absorption region exhibits an age gradient similar to the global trend, wherein the older population (4 Gyr) in the east transitions to a younger population in the west. The strongest absorption features are related to the youngest stellar populations ($\sim 1.4$ Gyr), which is an indicator of the last starburst episode in the highlighted region. Lack of measurable H$\beta$ emission lines indicates no detectable ongoing star-formation \citep{2018MNRAS.477.1708P}; indeed, the spatially resolved BPT diagram only shows signatures of star formation on the western edge of the contoured Balmer absorption region, where the stellar populations of Mrk 739 reach their minimum age with values between $\sim 0.8$ to 1 Gyr.

The kinematics of the stellar populations in the system present in general velocity dispersion values close to the spectral resolution of MUSE at the north and south of the galaxy. The regions associated with the nuclei present higher velocity dispersion, namely $\sim$90 km s$^{-1}$ for Mrk 739W and up to $\sim$120 km s$^{-1}$ in Mrk 739E. We see that the velocity dispersion is near $\sim$140~km s$^{-1}$ at the East of Mrk 739E, and it is not spatially peaked with the center of the galaxy, which is assumed to be at the location of the AGN. The velocity map shows a clear gradient of $\sim60$ km s$^{-1}$ from north to south, and an old-population on the southeast with velocities of $\sim -100$ km s$^{-1}$. The bins of the velocity map surrounding the East nucleus show a velocity gradient of $\Delta V \sim 100$ km s$^{-1}$ in the west to east direction, but it is not a decoupled structure with an evident inclination as in the ionized gas disk. The stellar velocity gradient has an amplitude three times lower than the velocity gradient present in the ionized gas kinematic and is oriented in the opposite direction. Interestingly, this suggests that there is not evidence of a rotating disk surrounding the East AGN, as is indeed present in the ionized gas, implying it was likely driven there relatively recently.

In Figure~\ref{fig:SPsn40padova} we presented luminosity-weighted average ages and metallicities. It is however instructive to study also the full distribution of stellar templates contributing to this fit. In order to do this, in Figure \ref{fig:agecontribution} we present the spatial distribution and fractional contributions to the best-fit of each stellar template, distributed in four stellar ages bins ($0-2$, $2-6$, $6-9$, and $9-14$ Gyr). Mrk 739W is strongly dominated by young and medium-age populations. For instance, almost 80\% of the stars in the west side of Mrk 739W range between $0-2$ Gyr. The medium age bin ($2-6$ Gyr) contributes $\sim 50\%$ of the light in the contoured Balmer absorption region, while older ($\sim 7$ Gyr) populations are also relevant, contributing with $\sim 30 -40\%$ to this region. Mrk 739E shows instead a lower ($\sim 15\%$) fraction of young stellar population and is mostly dominated by older populations, as can be seen particularly in the southeast region of the galaxy.

\subsection{Estimation of SFRs and stellar masses}\label{sfr_stellarmass}

According to \cite{2011ApJ...735L..42K}, the SDSS \textit{r}-band magnitudes obtained from two-dimensional surface brightness fitting  are $m_{r}=13.75\pm0.15$ for Mrk 739W and $m_{r}=14.03\pm0.15$ for Mrk 739E. Considering that the luminosity at longer wavelengths can be used as a tracer of the stellar mass of the galaxy \citep{2001MNRAS.326..255C,0067-0049-149-2-289}, the ratio between the \textit{r}-band magnitudes gives a rough estimation of the ratio between the stellar masses of each galaxy in Mrk 739. As a consequence, it can be concluded that the two galaxies have roughly the same mass, therefore qualifying their interaction as a major merger with $M/m<3$. 

In addition, the near-IR luminosity for example in the K-band (2.2$\mu$m) is a good tracer of the stellar mass, since most of the K-band light of the stellar population arises from long-lived giant stars and is less affected by dust obscuration than shorter wavelengths
\citep[e.g.,][]{10.1046/j.1365-8711.1998.01708.x}. The infrared K-band magnitudes of Mrk 739E and Mrk 739W, as reported by \citet{Imanishi_2013}, are $m_{K}=11.37$ and $m_{K}=12.74$, respectively. AGN emission weakly obscured by dust, as in Mrk 739E (see \S \ref{extinction}), can contribute substantially to the observed K-band flux \citep{Imanishi_2013}. Therefore, we subtracted the contribution of the inner $0\farcs5$ to the total K-band flux \citep[see Table 3 in][]{Imanishi_2013} in Mrk 739E, assuming that this region is fully dominated by AGN emission. The AGN-corrected K-band magnitude for Mrk 739E is then $m_{K}=11.83$. Following the discussion in \cite[][]{2013ARA&A..51..511K}, we derive stellar masses of $\log (M_{*}/M_{\odot})=10.86$ and $\log (M_{*}/M_{\odot})=10.50$ for Mrk 739E and Mrk 739W, respectively. The K-band yields a more accurate estimation of the stellar mass ratio, where $M_{\rm East}/M_{\rm West}=2.2$. Mrk 739W is brighter than Mrk 739E in the \textit{r}-band since the young stellar populations ($1-2$ Gyr), located at the West direction of the system, are expected to be hot, have short lifetimes, and contribute more at short wavelengths. The oldest stellar populations in Mrk 739E confirm the fact that this galaxy is more luminous at redder wavelengths, particularly in the K-band. 

The SFR derived by adopting the \cite{Kennicutt_1998} relation, where $\rm SFR (M_\odot/\rm yr)=7.9\times 10^{-42}L_{H_{\alpha}} (\rm erg \; \rm s^{-1})$, for Mrk 739W is $\rm SFR_{H\alpha} (M_\odot/\rm yr)=5.3$. This relation considers a Salpeter IMF, while the luminosity of the H$\alpha$ was corrected by stellar absorption and dust extinction based on the Balmer decrement $\rm H\alpha/\rm H\beta$ flux ratio, asp presented in section 5.1. We further assume that all the H$\alpha$ emission at the West of Mrk 739E is associated with star formation and belong to Mrk 739W. The SFR of Mrk 739W, based on the UV emission reported by \cite{2011ApJ...735L..42K}, is $\rm SFR_{\rm UV} (M_\odot/\rm yr)=0.6$, while the unresolved SFR for Mrk 739 derived from far-infrared (FIR) is $\rm SFR_{\rm FIR} (M_\odot/\rm yr)=6.9$. SFR measurements on Mrk 739E are very unreliable due to AGN contamination. However, according to our optical diagnostic diagram, there are not significant star-forming regions related to the eastern galaxy. Thus, Mrk 739W would lie, within the scatter, on the star-forming main sequence of galaxies \citep[e.g., see][]{2010ApJ...721..193P,Renzini_2015}, consistent with ongoing star formation and low AGN activity. Mrk 739E, on the other hand, could be undergoing significant star formation quenching, likely related to AGN activity \citep{McPartland_2018}, without predominant star-forming regions and old stellar populations.  

\section{Discussion}\label{discuss}

\subsection{Extinction map}\label{extinction}

\begin{figure}
\includegraphics[width=1.0\columnwidth]{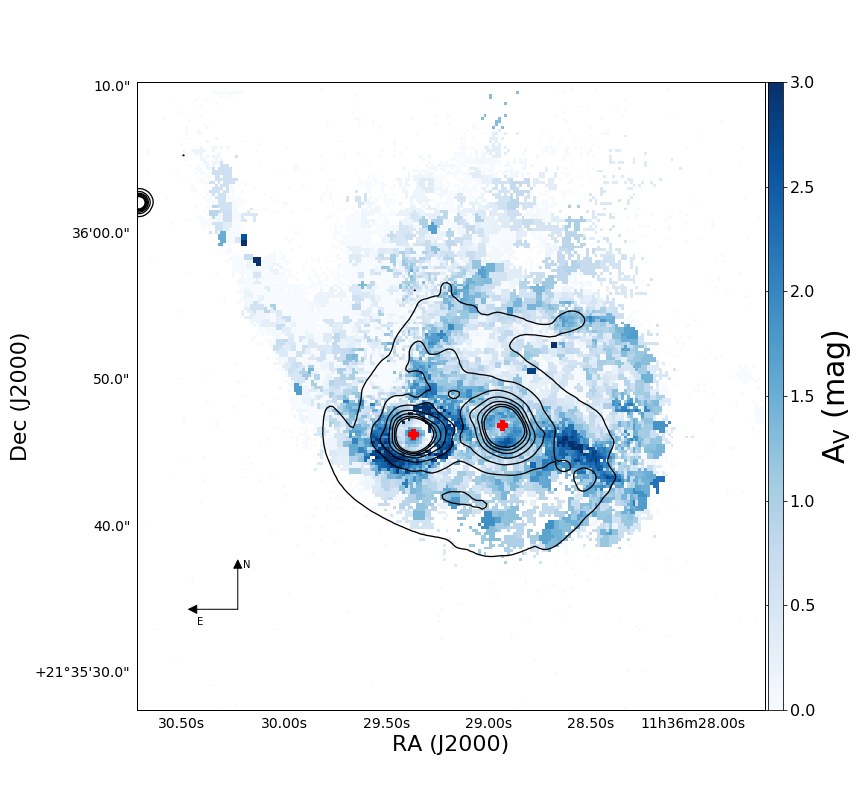}
\caption{Optical extinction A$_{\rm V}$ map produced by dust attenuation, as computed using the Balmer decrement $\rm H\alpha/\rm H\beta$ flux ratio. Red crosses mark the positions of the central AGNs. The contours of the white-light image presented in Figure \ref{fig:whiteimage} are overlaid.}\label{OE}
\end{figure}

The H$\alpha$ to H$\beta$ flux ratio can be used to estimate the extinction by dust in the line of sight, as it was proposed by \citet{1986A&A...155..297C} based on observations of HII regions in the Large Magellanic Cloud. 
In Figure \ref{OE}, we report the total extinction, A$_{\rm V}$, in the $V$ band, based on the Balmer decrement H$\alpha/$\rm H$\beta$ flux ratio. Following the procedure of \cite{Dom_nguez_2013}, we considered the reddening curve proposed by \cite{2000ApJ...533..682C}, an intrinsic Balmer ratio of $(\rm H\alpha/\rm H\beta)_{int}=2.86$ (for an electron temperature T$_{e}$= $10^4$K;  \citealt{2006agna.book.....O}) adopting case B recombination, a Milky Way extinction curve and an attenuation law for galactic diffuse ISM $R_{V} =3.12$ \citep{2000ApJ...533..682C}. 

The map shows signatures for relatively high extinction surrounding both nuclei, and in particular, close to Mrk 739E, reaching extinction values $\rm A_{\rm V}\sim$3 at the location of the rotating disk. We also observe high values of dust extinction associated with the optically-diagnosed star-forming regions described in \S \ref{subsec:natureofionization}, while most of the galaxy presents more moderate optical extinction between $\sim 0-2$. The extinction map reveals an extended arc-shaped distribution reaching slightly higher values than its surrounding regions. Starting at the North of the East nucleus, the dust extinction, with values A$_{\rm V}\sim 1.5$, seems to be defining the perimeter of the leading edge of the potential spiral arms of Mrk 739W, which can just barely be seen in the white-light image in Figure \ref{fig:whiteimage}. To the West direction, the distribution follows a semi-circular shape that ends where our optical diagnostic diagrams revealed an extended star-forming region.

\subsection{Super Massive Black Holes masses}\label{smbhmass}

A variety of techniques, such as dynamical, spectral fitting, and scaling methods can be used to estimate the mass of the central SMBH in galactic nuclei \citep{2010MmSAI..81..281C,2013BASI...41...61S}. For instance, one way to compute this value in AGN is to use scaling relations linking the BLR properties, the optical continuum flux, and specific emission lines through scaling factors \citep{2002ApJ...571..733V,2012ApJ...753..125S,2012NewAR..56...49M}. Specifically, \cite{2000ApJ...533..631K} reported that the BLR size scales with the $5100$\AA~luminosity as R$_{\rm BLR}$ $\propto L^{0.70\pm0.3}$. Following this procedure, \cite{2006ApJ...641..689V} presented four empirical relations to estimate the central black hole mass in nearby AGNs based on single-epoch spectra. Specifically, we use equation 6 from their work:

\begin{multline}
    \log M_{\rm BH}/M_{\odot}(\rm H\beta)=\\\log \left\lbrace\Bigg[\frac{FWHM(\rm H\beta)}{1000\;\rm km\;\rm s^{-1}}\Bigg]^{2}\Bigg[\frac{L(\rm H\beta)}{10^{42}\;\rm erg\;\rm s^{-1}}\Bigg]^{0.63}\right\rbrace\\ + (6.67\;\pm\;0.03).
\end{multline}\label{masshbeta}
\noindent 
From a spectrum extracted in a region centered on the Mrk 739E nucleus, with a radius of $0\farcs9$ we obtain values of FWHM$(\rm H\beta)$=4285$\pm$382 km s$^{-1}$ and a luminosity of log L(H$\beta$) [$\rm erg \;\rm s^{-1}$]=40.87$\pm$0.28, which leads to a derived SMBH mass of $\log M_{\rm BH}/M_{\odot}$=7.22$\pm$0.25. 
This is slightly higher, but consistent, with the value found by
\cite{2011ApJ...735L..42K} of $\log M_{\rm BH}/M_{\odot}$=7.04$\pm$0.40, which is
based on the FWHM(H$\beta$)-L$_{\lambda}(5100$ \AA$)$ scaling relation from
\cite{2006ApJ...641..689V}. 

We can compare the value of the SMBH mass in Mrk 739E with the value obtained from the M$_{BH}$-$\sigma_{*}$ relation \citep{2000ApJ...539L...9F,2000ApJ...539L..13G,2009ApJ...698..198G,2013ARA&A..51..511K}. From the stellar population maps presented in \S \ref{stellarpopulation},  we estimate the velocity dispersion of the spheroidal component by averaging the bins surrounding the masked region of the East nucleus within an aperture of $0\farcs9$, obtaining a value of $\sigma_{*}$=114$\pm$10~km s$^{-1}$. Then, following equation 3 in \citet{2009ApJ...698..198G} we estimate a SMBH mass of $\log\;M_{\rm BH}/M_{\odot}$=7.09$\pm$0.15, fully consistent with the value derived from the H$\beta$ broad emission line. This consistency indicates that the East nucleus is located, within the scatter, on the $M_{\rm BH}-\sigma_{*}$ relation derived for normal and non-merging galaxies.

The SMBH in Mrk 739E is, on average, almost two orders of magnitude less massive than the typical values of 9 nearby (ultra-)luminous infrared late-stage merging galaxies reported by \citet{2015ApJ...803...61M}. Furthermore, they found that their sample of black holes does not follow the $M_{\rm BH}-\sigma_{*}$ relation given by \cite{2013ApJ...764..184M} and the SMBH sample are overmassive compared to the expected values based on scaling relations, suggesting that the major epoch of black hole growth occurs in the early stages of the merger.
Since the SMBH in Mrk 739E lies on the $M_{\rm BH}-\sigma_{*}$ relation, we can speculate that it, in a similar way as the SMBHs reported in \citet{2015ApJ...803...61M}, could grow most of its mass during the collision due to this assembly delay that allows mergers to move off the relation, departing from the $M_{\rm BH}-\sigma_{*}$. This scenario supports the idea that the Mrk 739 system is in an early evolutionary stage, and the SMBHs have not started yet to grow due to the merger. To move off up to 3-sigma of the relation, the SMBH in Mrk 739E would need to grow by two orders of magnitude, requiring large ($>10^{8}\rm M_{\odot}$) amounts of gas to be funneled down to an accretion disk, together with Eddington-limited accretion for a few tens to hundreds of millions of years \citep{2015ApJ...803...61M}. High-resolution observations of the molecular gas can lead us to resolve the amount of available gas that could be accreted onto the SMBHs, and thus understand the extent to which this might be possible, estimating the eventual values that can be reached by the SMBH mass in Mrk 739E.

The sphere of influence of a black hole is defined as $r_{infl}=GM_{\rm BH}/\sigma_{*}^{2}$ \citep{2004cbhg.symp..263M}, where $G$ is the gravitational constant and $\sigma_{*}^{2}$ is the velocity dispersion of the stars of the bulge. Since the East SMBH has a mass of $\log M_{\rm BH}/M_{\odot}$=7.22$\pm$ 0.25, the sphere of influence of the East SMBH is $r_{infl}=5.57\pm2.80$ pc, which is unresolved with our MUSE observation. These scales are significantly ($\sim 3$ orders of magnitude) smaller than the mass and radius of the rotating disk, and hence we conclude that the latter is mostly unaffected by the gravitational field of the SMBH at the center.

The mass of the SMBH in the West nucleus cannot be determined using our MUSE data and the scaling relation from equation \ref{masshbeta}. It is because the broad Balmer lines are not visible, likely due to a combination of obscuration and dilution by the host galaxy ({\it lower-right} panel in Figure \ref{fig:representativespectra}). Based on the X-ray luminosity of Mrk 739W, $L_{2-10 \rm keV}=1.0\times10^{42}$ erg s$^{-1}$, the expected extinction-corrected H$\alpha$ luminosity, mostly from the BLR, is $L_{\rm H_{\alpha}}\sim1\times10^{41}\; \rm erg \rm s^{-1}$, via \citet{2006A&A...455..173P} relation. With an extinction of $\rm A_{V} \sim 1-2$, the expected observed luminosity of the BLR in Mrk 739W would be $L_{\rm H_{\alpha}}\sim1.6-4\times10^{40}\; \rm erg \rm s^{-1}$. Since H$\alpha$ emission in Mrk 739W is extended and slightly more luminous ($L_{\rm H_{\alpha}}\sim6.2\times10^{40}\; \rm erg \rm s^{-1}$), we expect a indistinguishable BLR given the extinction reported in \S \ref{extinction}. Higher angular resolution spectroscopy, e.g., from MUSE NFM observations, might resolve the extended narrow line and better recover any possible point-like broad component. Also, it would be useful to obtain near-IR spectroscopy to study the hydrogen Paschen series transitions \citep[e.g.,][]{landt13,lafranca15} and test whether there is a hidden broad-line region, though these are rare \citep[e.g.,][]{2017MNRAS.467..540L,onori17}. Given the fact that Mrk739W is optically obscured but shows X-ray emission, the X-ray luminosity along with near-infrared emission lines might also be a promising path to measure the SMBH mass \citep{2017A&A...598A..51R}. Assuming that the Mrk 739W SMBH, as Mrk 739E, lies on the $M_{\rm BH}-\sigma_{*}$ relation, we can obtain a rough estimate of its mass.  With an average stellar velocity dispersion of $\sigma_{*} = 91 \pm 5 $ km s$^{-1}$ within an aperture of $0\farcs9$, we obtain a SMBH mass for the West nucleus of $\log M_{\rm BH}/M_{\odot}$=6.68$\pm$0.045, adopting the $M_{\rm BH}-\sigma_{*}$ relation of \citet{2009ApJ...698..198G}. 

For a bolometric luminosity of $1\times10^{45}\rm erg \;\rm s^{-1}$ and a SMBH mass of  $\log M_{\rm BH}/M_{\odot}$= 7.04$\pm$0.4, \cite{2011ApJ...735L..42K} found that the Eddington ratio, defined as $\lambda_{Edd}= L_{bol}/L_{Edd}$\footnote{$L_{Edd}$ is the Eddington luminosity defined as $L_{Edd}=4\pi c G M_{BH} m_{H}/\sigma_{T}$}, is $\lambda_{Edd}=0.71$ for Mrk 739E. For the SMBH mass value of $\log M_{\rm BH}/M_{\odot}$= 7.22$\pm$0.25 that we derived, we estimate a $\lambda_{Edd}=0.48$. Assuming that the SMBH mass of Mrk 739W obtained from the $M_{\rm BH}-\sigma_{*}$ relation is correct, the Eddington ratio for Mrk 739W is $\lambda_{Edd}=0.033$ for a bolometric luminosity of $2\times10^{43}\rm erg \;\rm s^{-1}$ derived in  \citet{2011ApJ...735L..42K}. Hence, the SMBHs seems to be in different accretion scenarios. While our measurement of the Eddington ratio of Mrk 739E is slightly lower than the value reported by \citet{2011ApJ...735L..42K}, it is still one of the highest ratios among the \textit{Swift} BAT AGN \citep{2010MNRAS.402.1081V}. On the other hand, the Eddington ratio of Mrk 739W is consistent with the median $\lambda_{Edd}$ value of the \textit{Swift}-BAT AGN Spectroscopic Survey (BASS) \citep{Koss_2017}. High resolution infrared imaging in K- (2.2$\mu$m), L-bands (3.8$\mu$m), and red $K-L$ colors, sensitive to buried AGN \citep{Imanishi_2013}, could not find evidence for an AGN in Mrk 739W, suggesting non-synchronous mass accretion onto SMBHs \citep{2012ApJ...748L...7V}, consistent with the Eddington ratio estimations for Mrk 739E and Mrk 739W. 

\subsection{Morphology and possible evolution of the system }\label{evolution}

\begin{figure*}
\includegraphics[width=1.0\textwidth,keepaspectratio=true]{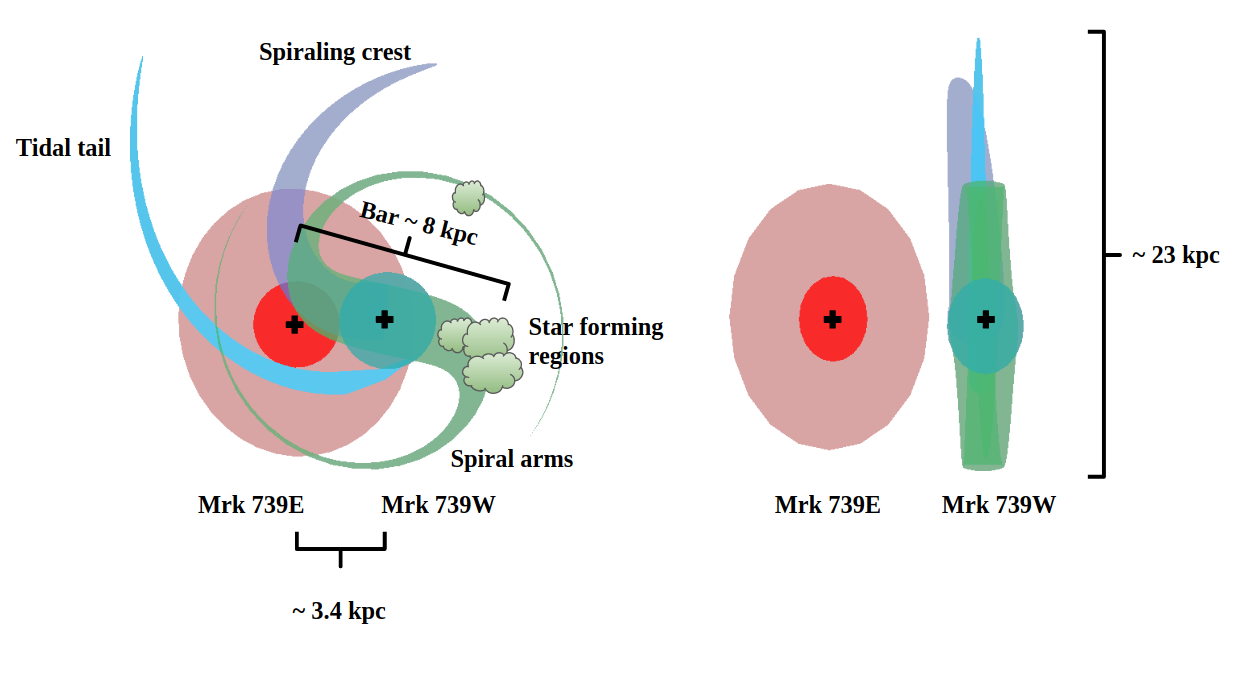}
\caption{Schematic representation of the dual AGN Mrk 739 as described in \S \ref{evolution}. {\it Left panel}: Face-on view of the galaxy oriented North up and East to the left. {\it Right panel}: Edge-on view from the East. Mrk 739W is assumed to be a spiral-barred galaxy ({\it green}) with an elongated tidal tail ({\it light-blue}), and the spiral arm ({\it blue}) that gives origin to the spiraling crest. The clouds ({\it green}) mark the location of the star-forming regions described in \S \ref{subsec:natureofionization}. Mrk 739E is represented as an elliptical galaxy ({\it red}). The bulges are shown as spheres at the centers of Mrk 739E and Mrk 739W ({\it red} and {\it green} respectively), while the black crosses mark the location of the AGN. The scales correspond to the projected distances of the bar, the separation between the nuclei, and the size of the entire system in the north-to-south direction. The distance between Mrk 739E and Mrk 739W in the {\it right panel} is arbitrary for visualization purpose.}  
\label{fig:squematic}
\end{figure*}

The VLT/MUSE observations presented here can provide valuable insights into the current merger stage of the dual AGN, Mrk 739. Morphologically, we present a schematic illustration of the dual AGN in Figure \ref{fig:squematic}. Mrk 739W resembles a spiral-barred galaxy, with spiral arms that spatially overlap with Mrk 739E. This overlap likely explains the double narrow-line component region ({\it top-right} panel in Figure \ref{fig:representativespectra}), where the redshifted emission line component is likely associated with Mrk 739W's bar, and the blueshifted one would come from Mrk 739E's disk region. The eastern side of the bar is the origin of two spiral arms, one of them extending up to the western star-forming region, while the other is morphologically related to the spatially-extended region to the north side of Mrk 739, the so-called spiraling crest. The southern arm extends up to the eastern side of Mrk 739E. 

Dynamically, the common velocity between the eastern tidal tail ({\it light-blue} in Figure \ref{fig:squematic}) and Mrk 739W in the ionized gas suggests that the tail can be associated with the West galaxy rather than to the East one. The spiraling crest with velocities of $\sim$60 km s$^{-1}$ (Figure \ref{fig:kinematics}) is associated with the leading arm of the northern spiral arms of Mrk 739W. The stellar kinematics suggest that Mrk 739W is mostly face-on, with a stellar velocity gradient of only $\sim$60 km s$^{-1}$ from the north-to-south direction ({\it lower-left} panel in Figure \ref{fig:SPsn40padova}). 

Since we only distinguish Mrk 739E by its rotating disk of ionized gas and its old, blueshifted, and metal-poor population on the southeast edge of the galaxy, we claim that the young stellar populations of Mrk 739W are hiding and extinguishing the East galaxy from a foreground position, making Mrk 739E nearly absent on luminosity-weighted maps close to the center of the system, and highly extincted with values of $\rm A_{\rm V}\sim 3$. The high stellar velocity dispersion region on the east side of the eastern nucleus thus represents an interface region where both galaxies appear in projection.

The morphology and dynamics of the dual AGN Mrk 739 are hence consistent with an early stage of the collision, where the foreground galaxy is a young star-forming galaxy that is interacting with its background elliptical companion. Since Mrk 739W's AGN does not show evidence of being actively accreting, as we discussed in \S \ref{smbhmass}, we propose that the nuclear activity of Mrk 739E has been the main ionizing mechanism of the northwestern spiral arms, similar to the phenomenon known as ``Hanny's Voorwerps'', as seen for example in IC 2497 \citep[see, e.g.,][]{lintott_2009,jozaJ,Keel_2012,Sartori_2016}. This scenario suggests that the East AGN ionized the northern structures of Mrk 739W from its background-position.

We emphasize the need for high-resolution multi-wavelength and spatially-resolved spectroscopic data, covering optical, near-IR, and sub-mm wavelengths, to dynamically model the system in order to understand the exact evolutionary stage of this complex merging galaxy. In particular, this dataset would allow characterizing the behavior of the gas that is actively feeding both AGNs, the distribution of the dense and cold gas close to the nuclei that work as a reservoir for future accretion onto both SMBHs, the stellar structures that could be produced by the interaction, and stellar formation related to the West galaxy.

\section{Conclusions}

We have carried out a comprehensive morphological and kinematic study of the nearby dual AGN Mrk 739 using VLT/MUSE Wide Field Mode observations. We studied the optical emission lines to map the behavior of the ionized gas in the galaxy, revealing an extended and intense northern spiraling crest at the North of both nuclei, which does not has a symmetric equivalent to the south, mainly in the [OIII]$\lambda5007$ emission line map. The [OIII] line displays a non-clumpy distribution, a visible tidal tail ascribable to the ongoing major merger, and a lack of emission to the south of the nuclei.  H$\alpha$ and H$\beta$ maps reveal a different spatial distribution compared to the [OIII] line. The Balmer lines were found more prominently near both nuclei and present slightly more irregular structures, such as those detected at the West and North-West direction of the galaxy in Figures \ref{fig:halphahbetaoiiioi} and \ref{fig:NIIbSII}. The presence of Balmer absorption features is related to the weak or complete absence of ionized gas and post-starburst episodes, probably linked with the major merger. We conclude that the spiraling crest at the north direction of Mrk 739E is associated with the leading spiral arms originated at the north side of the bar. We found that the redshifted line of the double-peaked emission line region is also originated by the presence of the bar, while the blueshifted line is related to the rotating disk that surrounds the eastern nucleus with a circular velocity of $237^{+26}_{-28}$ km s$^{-1}$, an inclination of $33^{+5}_{-3}$ degrees, and with a dynamical mass of $\log M(M_{\odot})=10.20\pm0.06$, $\sim 1000\times$ larger than the SMBH mass of Mrk 739E. Since Mrk 739E is an elliptical galaxy and its stellar kinematics does not show evidence of a kinematically decoupled structure, we claim that the rotating disk of ionized gas was driven there relatively recently. Probably the rotating disk is not in dynamical equilibrium yet, and it is formed by the remnant ionized gas of the elliptical galaxy that was driven toward the center due to the gravitational interaction, explaining the high dynamical mass found.

From the spatially-resolved emission-line ratios and BPT diagrams, we studied the nature of the source of ionizing radiation. The galaxy has an extended AGN-ionized emission-line region that covers several ($5-20$) kpc from the nuclei. The star-forming regions, traced by areas where the H$\alpha$ emission line map is higher than [NII] and [SII] lines, are confined ($2-3$ kpc) around the nuclei, consistent with what \cite{1978Afz....14...69P}, \cite{1987A&A...171...41N}, and \cite{2011ApJ...735L..42K} have found previously. Since there is evidence of low accretion rates, an optically obscured AGN, and low X-ray luminosity in the western nucleus, we claim that the AGN-ionized emission-line region at the north of Mrk 739W is directly triggered by the AGN activity of Mrk 739E. This scenario suggests that Mrk 739E follow a northwest-to-southeast trajectory, ionizing the northern structures of Mrk 739W before its current position.

We found a gradient of $\sim 60$ km s$^{-1}$ from north-to-south on the stellar velocity map that characterizes the stellar orbit of the face-on main disk in Mrk 739W. The velocity dispersion map shows higher velocities close to the nuclei and an interface region with $\sigma_{*} = 140$ km s$^{-1}$. The blueshifted region at the southeast of the stellar population maps reveals an old population, suggesting that Mrk 739E is an old elliptical galaxy located behind Mrk 739W. 

Mrk 739 stands as a unique system that hosts two different accretion scenarios for its SMBHs. Mrk 739E host a rapidly accreting SMBH that lies on the $M_{\rm BH}-\sigma_{*}$ relation, while Mrk 739W presents evidence of an AGN obscured by star formation with a low-Eddington ratio that suggests non-synchronous mass accretion onto SMBHs. Mrk 739W appears to lie on the star-forming main sequence, with a $\rm SFR_{H\alpha} (M_\odot/\rm yr)=5.3$, while Mrk 739E shows signs of SF quenching. This scenario is similar in terms of star formation, to the nearby dual AGN Mrk 463, where values of SFR $\sim30$ and $\sim0.75$ M$_\odot/\rm yr$ were found for Mrk 463E and Mrk 463W, respectively \citep{2018ApJ...854...83T}. On the other hand, the nearby dual AGN NGC 6240 reach SFR values of $\sim100$ M$_\odot/\rm yr$ \citep{M_ller_S_nchez_2018}, illustrating the highly complex and dynamic features of a dual AGN. In terms of the evolutionary stage, unlike Mrk 463 and NGC 6240, which are at a more advanced merger stage, Mrk 739 is likely in a first-encounter phase, where both galaxies are beginning the galactic collision.

The need for higher spatial resolution of the order of 10s-1000s pc in optical IFU observations with AO, such as those provided by VLT/MUSE in its Narrow Field Mode, will be crucial to disentangle the distribution of the gas closer to the SMBHs. Multiwavelength data capable of match the high spatial resolution of MUSE in AO, for example, ALMA data, would be crucial to characterize the molecular gas which is actually forming stars in the West nucleus and feeding the AGN in the East galaxy, thus, along with data from others dual AGN, offer a better understanding of how the gas, dust, and stars behave during major galaxy mergers and the simultaneous activation of the nuclear sources in the host galaxies.  

\acknowledgments 

We thank the anonymous referee for his/her very useful and constructive suggestions. DT, ET, GDA, FB, and GV acknowledge support from CATA-Basal AFB-170002; DT, ET, and FB acknowledge support from FONDECYT Regular grant 1190818, GV acknowledge support from ANID program FONDECYT Postdoctorado 3200802, MK acknowledges support from NASA through ADAP award 80NSSC19K0749, FMS acknowledges support from NASA through ADAP award:  80NSSC19K1096, while ET further acknowledges support from ANID Anillo ACT172033 and Millennium Nucleus NCN19\_058 (TITANs). This work is based on observations collected at the European Southern Observatory under ESO program 095.B-0482. This research has made use of data obtained from the Chandra Source Catalog, provided by the Chandra X-ray Center (CXC) as part of the Chandra Data Archive. The Geryon cluster at the Centro de Astro-Ingenieria UC was extensively used for the calculations performed in this paper. BASAL CATA PFB-06, the Anillo ACT-86, FONDEQUIP AIC-57, and QUIMAL 130008 provided funding for several improvements to the Geryon cluster.

\software{ESO VLT/MUSE pipeline \citep{2014ASPC..485..451W}, ESO \textit{Reflex} environment \citep{2013A&A...559A..96F}, Pyspeckit \citep{2011ascl.soft09001G}, pPXF \citep{Cappellari2017}, KinMS \citep{2013MNRAS.429..534D}}

\bibliography{biblio.bib}


\end{document}